\documentclass[twocolumn]{aastex631}

\usepackage{graphicx} 

\graphicspath{{figures/}}

\begin{document}

\title{StarFlow: Leveraging Normalizing Flows for Stellar Age Estimation in SDSS-V DR19}
\author[0000-0003-4761-9305]{Alexander Stone-Martinez }
\affiliation{Department of Astronomy, New Mexico State University, P.O.Box 30001, MSC 4500, Las Cruces, NM, 88033, USA}

\author[0000-0002-9771-9622]{Jon A. Holtzman}
\affiliation{Department of Astronomy, New Mexico State University, P.O.Box 30001, MSC 4500, Las Cruces, NM, 88033, USA}

\author[0000-0003-4769-3273]{Yuxi(Lucy) Lu}
\affiliation{Department of Astronomy, The Ohio State University, Columbus, 140 W 18th Ave, OH 43210, USA}
\affiliation{Center for Cosmology and Astroparticle Physics (CCAPP), The Ohio State University, 191 W. Woodruff Ave., Columbus, OH 43210, USA}

\author[0000-0001-5388-0994]{Sten Hasselquist}
\affiliation{Space Telescope Science Institute, 3700 San Martin Drive, Baltimore, MD 21218, USA}

\author[0000-0003-2025-3585]{Julie Imig}
\affiliation{Space Telescope Science Institute, 3700 San Martin Drive, Baltimore, MD 21218, USA}

\author[0000-0001-9345-9977]{Emily J. Griffith}
\altaffiliation{NSF Astronomy and Astrophysics Postdoctoral Fellow}
\affiliation{Center for Astrophysics and Space Astronomy, Department of Astrophysical and Planetary Sciences, University  of Colorado, 389~UCB, Boulder,~CO 80309-0389, USA}

\author[0000-0003-4456-4863]{Earl P. Bellinger}
\affiliation{Department of Astronomy, Yale University, New Haven, CT 06510 USA}

\author[0000-0002-6561-9002]{Andrew K. Saydjari}
\altaffiliation{Hubble Fellow}
\affiliation{Department of Astrophysical Sciences, Princeton University,
Princeton, NJ 08544 USA}


\begin{abstract}

Understanding the ages of stars is crucial for unraveling the formation history and evolution of our Galaxy. Traditional methods for estimating stellar ages from spectroscopic data often struggle with providing appropriate uncertainty estimations and are severely constrained by the parameter space. In this work, we introduce a new approach using normalizing flows—a type of deep generative model—to estimate stellar ages for evolved stars with improved accuracy and robust uncertainty characterization. The model is trained on stellar masses for evolved stars derived from asteroseismology and predicts the relationship between the carbon and nitrogen abundances of a given star and its age. Unlike standard neural network techniques, normalizing flows enable the recovery of full likelihood distributions for individual stellar ages, offering a richer and more informative perspective on uncertainties. Our method yields age estimations for 378,720 evolved stars and achieves a typical absolute age uncertainty of approximately 2 Gyr. By intrinsically accounting for the coverage and density of the training data, our model ensures that the resulting uncertainties reflect both the inherent noise in the data and the completeness of the sampled parameter space. Applying this method to data from the SDSS-V Milky Way Mapper, we have produced the largest stellar age catalog for evolved stars to date.

\end{abstract}

\section{Introduction} \label{sec:intro}

Understanding the ages of stars is crucial for unraveling the formation history and evolution of our Galaxy. The Milky Way offers a unique opportunity to study galactic formation up close by resolving individual stars and directly measuring their properties. While extragalactic studies beyond the local group rely on integrated light and population synthesis models, the detailed observations within our own Galaxy allow for a more precise reconstruction of its chemodynamical evolution. Spectroscopic surveys such as SDSS-IV/APOGEE (\citealp[Apache Point Observatory Galactic Evolution Experiment;][]{APOGEE2017,APOGEE1_2019PASP}), SDSS-V/MWM (\citealp[Milky Way Mapper;][]{BOSS_2013AJ,SDSS52019BAAS}, Kollmeier et al. (2025) in preparation), GALAH (\citealp[GALactic Archaeology with HERMES;][]{Buder2018GALAH, deSilvaGalah}), LAMOST (\citealp[Large Sky Area Multi-Object Fiber Spectroscopic Telescope;][]{lamost2012,dengLamost}), and Gaia-ESO \citep{GaiaESO2012}, along with astrometric data from \textit{Gaia} \citep{Gaia2016}, have provided unprecedented insights into the chemical abundances, kinematics, and dynamics of millions of stars. However, accurately determining stellar ages remains one of the most challenging tasks in astrophysics because age is not a directly observable property \citep[e.g.,][]{soderblom2010}.

Determining stellar ages is especially challenging because it must be inferred from other measurable quantities, leading to degeneracies and uncertainties. Traditional methods like isochrone fitting compare observed stellar parameters—such as effective temperature and luminosity—to theoretical models to estimate ages (\citealp{Pont2004, Dotter2008, Breean2012, Serenelli2013Iso}). This technique is effective for certain evolutionary phases, such as stars on the sub-giant branch, where evolutionary tracks in the Hertzsprung-Russell diagram are well-separated. However, for more evolved stars, evolutionary tracks converge, resulting in degenerate solutions and increased uncertainties in age estimation. Observational uncertainties further compound these difficulties, making precise age determination problematic. 

Asteroseismology, which measures oscillations within stars, provides highly precise age estimates for red giant stars, with uncertainties around 10\% and down to 1\% for some sub-giants \citep{2017Miglio,Garcia2019Seismo}. This can be accomplished with grid-based modeling (\citealp{2013Chaplin,Aguirre2015, Aguirre2017, Aguirre2018}), or by employing a series of scaling relations that link asteroseismic properties, such as \(\Delta \nu\) and \(\nu_{\text{max}}\), to stellar mass \citep{BellingerScaleing2020}. Following this, the derived mass estimates are applied to a set of models to produce age estimations. Missions like \textit{Kepler}, K2, TESS, and the upcoming PLATO mission(\citealp{Borucki2010TESS, Ricker2015Tess, 2017Miglio, 2025Plato}) have enabled precise asteroseismic measurements, but these data are limited to nearby stars in specific fields and require long time series observations, restricting their applicability to a small subset of stars which are insufficient for investigating large spatial variations of stellar properties across the Galaxy.

An alternative method involves using chemical clocks. The premise is that certain elemental abundances trace stellar age, usually [Fe/H], [C/N], [$\alpha$/Fe], or other elements such as R-process elements. [$\alpha$/Fe], [Fe/H] and R-process abundance based ages are predicted in classical chemical evolution models \citealp[e.g.][]{Matteucci2001,2009Pagel,nissen}, and observed in the solar neighborhood (\citealp{Fuhrmann2011, Haywood2013}). However, using ages estimated from these elements can introduce biases when examining chemical evolution due their calibration being dependent on current understanding of Galactic chemical evolution. In contrast, [C/N] is tied directly to stellar evolution (\citealp{Masseron_Gilmore2015,Martig2016, Roberts_2024MNRASR}).
The carbon-to-nitrogen abundance ratio ([C/N]) changes during the red giant branch phase due to internal mixing processes. As stars ascend the giant branch, their convective layers dredge up material processed in the carbon-nitrogen-oxygen (CNO) cycle, altering the surface [C/N] ratio in a way that correlates with stellar mass—and thus age \citep{Martig2016}. However, modeling these mixing processes accurately is complex, and uncertainties in stellar models make it difficult to predict [C/N] ratios precisely (\citealp{Masseron_Gilmore2015, Roberts_2024MNRASR}). Furthermore, stars with \([\rm{Fe/H}] < -0.5\) experience extra-mixing processes, complicating the [C/N]-age relationship \citep{ShetroneMixing}. These limitations make empirical calibration, such as using asteroseismic masses to calibrate the [C/N]-mass relation, an attractive option.

Machine learning (ML) has emerged as a powerful tool in spectroscopic age determination, offering a fast and powerful empirical approach. Studies have utilized ML models, often artificial neural networks (ANN), to find empirical relations between stellar parameters—including [C/N] abundances—and stellar ages estimated from asteroseismology (\citealp{Ness2016CANNON, Bellinger2016, MackerethAges, Hon2020, AndersAges, Leung_2023MNRAS, DistMass}). ML excels at capturing complex, nonlinear relationships within large and multi-faceted datasets, capable of integrating a broader range of stellar characteristics into its predictions more efficiently compared to conventional isochrone fitting for evolved stars. However, traditional ANNs have key limitations: they often lack robust uncertainty estimation, making it difficult to quantify the confidence in their predictions \citep{DistMass}. Moreover, they are prone to overfitting and lack the capability to store information about the training data's underlying distribution, which hinders their ability to assess data sparsity or how out of distribution a given data sample is. This makes them akin to sophisticated interpolation tools that may not generalize well beyond the training set.

To overcome these limitations, we turn to generative models, specifically normalizing flows—a class of ML techniques that can model complex probability distributions and provide robust uncertainty estimates. Generative models learn the underlying probability distributions of the data, allowing them to capture the full range of variability and uncertainties inherent in the observations. Normalizing flows, in particular, are powerful because they combine the flexibility of deep learning with exact likelihood estimation, making them well-suited for modeling complex, multimodal distributions (\citealp{normflowOverview,Weng_2018}).

For example, \citet{Leung_2023MNRAS} employed a type of generative model called a variational autoencoder (VAE) to estimate asteroseismic parameters from APOGEE spectra. While VAEs are effective in many scenarios and can mitigate overfitting by representing the latent space as a single Gaussian distribution, they may not adequately capture complex data distributions or model asymmetric uncertainties. Normalizing flows address these limitations by transforming simple probability distributions into complex ones through a series of invertible transformations, allowing for greater flexibility in modeling the data. Normalizing flows have been used in an astronomical context by \citet{Ting2022ApJ} to examine the relations between many elemental abundances and \citet{hon2024flowbasedgenerativeemulationgrids} to make a fast grid emulation tool.

In this work, we leverage normalizing flows to estimate stellar ages with improved accuracy and uncertainty characterization. By learning the full joint probability distribution of stellar parameters and abundances, our method provides robust age estimates while intrinsically accounting for the coverage and density of the training data. We utilize stellar parameters and abundances instead of the full APOGEE spectra. This is done since the parameters and abundance pipeline ASPCAP  utilizes the whole spectra to obtain the stellar parameters and abundances, so the input parameters of our model do not contain the signal noise found in the spectrum or any potential systematics such as LSF variation. This approach also facilitates directly examining the joint probability distribution to potentially improve our understanding of the parameter and abundance-age relation.

We validate our approach against asteroseismic data and cluster ages then apply it to the Milky Way Mapper DR19 catalog, producing a comprehensive stellar age catalog for galactic archaeology. This catalog will enable new insights into the formation and evolution of the Milky Way, contributing to our broader understanding of galactic dynamics and chemical evolution.

\section{Normalizing Flow Methodology}\label{sec:Method}
In general, we use normalizing flows to model the distribution of \(T_{\rm{eff}}\), \(\log g\), [Fe/H], [C/Fe], [N/Fe], and stellar age. Using the model distribution, we marginalize over all the parameters except stellar age to approximate the stellar age posterior for a given star.

 \subsection{Motivation for using normalizing flows}\label{sec:motivationMethod}
    
    Normalizing flows are particularly well-suited for our objectives, due to their ability to model complex, non-Gaussian distributions and to incorporate the reported uncertainties in the training data. 

    They have several advantages:
    \begin{enumerate}
        \item \textbf{Explicit Learning of Data Distribution:} Normalizing flows have the ability to explicitly learn the data distribution. This enables the model to capture complex distributions beyond simple Gaussian shapes. This capability is crucial for our work, as there is strong evidence suggesting that the [C/N] - Age relation breaks down at ages \( > 8\) Gyr and \( < 1\) Gyr \citep{Roberts_2024MNRASR}. This results in the age probability distribution being asymmetrical at the high age regime. Consequently, our reported uncertainties will encompass this asymmetry and thus will properly reflect the scatter in the training data.

        \item \textbf{Incorporation of Reported Uncertainties:} Including the reported uncertainties in the training data is a critical component of our motivation for using normalizing flows. This is done by sampling Gaussian distributions for each stellar parameter, centered on the reported value and with a width determined by the reported uncertainty. This approach allows us to make better use of the uncertainties associated with each parameter, and is implemented in both the training of the model and the extraction of the age probability distribution function (PDF).
        
        \item \textbf{Flexibility in Deriving Conditional PDFs:} We can derive the conditional probability distribution across any combination of parameters. Thus we can not only sample the mass PDF but also generate the 2D C-N PDF or the 3D C-N-Mass PDF. This capability enables direct examination of what the model predicts about the relationship between [C/N] and mass across parameter space. Additionally, this feature enables us to identify stars with atypical chemical compositions, which may lead to unreliable spectroscopic age estimates. We achieve this by detecting stars whose [C/Fe] and [N/Fe] abundances are highly improbable at any given mass, which may be due to variations in the initial abundances of C \& N when the star formed.

        \item \textbf{Density Information Retention:} Normalizing flows retain information about the overall distribution of the training data, including its density. This allows us to directly assess how well different regions of the parameter space are represented in the training dataset. Additionally, while other generative models such as VAEs have been shown to achieve accurate stellar ages from stellar spectra \citep{Leung_2023MNRAS}, they tend to be more constrained by the limitations in training set parameter space coverage. In contrast, the normalizing flow model can still fit a distribution to the sparsely covered regions of parameter space, which is particularly important for stars with lower surface gravity (\(\log g\)) since they are sparsely sampled in the training data.
        
        

    \end{enumerate}

    While normalizing flows effectively model complex, non-Gaussian distributions, they cannot distinguish between non-Gaussian features from astrophysical distributions and those from observational uncertainties. The model learns the joint distribution of stellar parameters and errors, capturing all variability as a unified probability density. As a result, non-Gaussian features could reflect either intrinsic stellar properties or measurement artifacts. This potential degeneracy does not compromise the model’s uncertainty estimates but complicates the interpretation of asymmetric or multimodal distributions. For stellar age estimation, this is unlikely to affect our results. However, caution is needed when using it to draw conclusions on the [C/N] - age relation itself.

    \subsection{Overview of Normalizing Flows}\label{sec:normflows}


     The core principle of a normalizing flow  is to map a random variable \(z\), which follows a simple probability distribution \(p_0(Z)\) (typically a multivariate Gaussian), to another variable \(x = f(z)\) that follows a more complex probability distribution \(p_k(x)\). The transformation function \(f = f_k \circ f_{k-1} \circ ... \circ f_{1}\) consists of a composition of functions, each with parameters that are tuned during model training. This transformation describes how \(z \approx p_0(z) \) maps to \( y \approx p_k(x)\), as governed by the change of variables theorem:
   \begin{displaymath}
    p_k(x) = p_{0}(z) \left|\det\left(\frac{\partial z}{\partial x}\right)\right|
    \end{displaymath}
    
   \begin{equation}
     = p_{0}(f^{-1}(z)) \left|\det\left(\frac{\partial f^{-1}(z)}{\partial z}\right)\right|
    \end{equation}

    \begin{equation}\label{flow}
    p_k(z_k) = p_0(z_0) \prod_{i=1}^{k} \left|\det\left(\frac{\partial f_i^{-1}(z_i)}{\partial z_i}\right)\right|
    \end{equation}

    Here, \( \det\left(\frac{\partial f^{-1}(z)}{\partial z}\right) \) is the determinant of the Jacobian matrix of the inverse transformation \( f^{-1} \). For this formulation to work, the transformation function \(f\)  must be both invertible and differentiable. It must be invertible so that \(x = f(z)\) and \(z = f^{-1}(x)\), and it must be differentiable to allow for the computation of the Jacobian determinant. By breaking down the transformation into each component function, we arrive at a multiplicative form, as shown in equation \ref{flow}. Each transformation \(f_i\) is referred to as a \textit{flow}, and the full chain of transformations constitutes a \textit{normalizing flow}.

    Training a normalizing flow model involves tuning the parameters of \(f\) to maximize
    \begin{equation}
    \log p(x) = \log p_k(z_k) = \log p_0(z_0) - \sum_{i=1}^{k} \log \left|\det\left(\frac{\partial f_i(z_{i-1})}{\partial z_{i}}\right)\right|
    \end{equation}
   This approach not only enables the modeling of complex, non-Gaussian distributions but also ensures that the likelihood of the observed data can be computed exactly. 
   
   An alternative approach to modeling complex distributions is to use K-process models or Gaussian Mixture Models (GMMs), which approximate distributions as a combination of multiple Gaussian components. This method effectively captures clusters or 'lumps' in the data, but struggles with continuous asymmetries, long tails, or complex correlations between variables. In contrast, normalizing flows transform a simple base distribution into a complex one using a series of flexible, invertible mappings. This allows them to model non-Gaussian features more generally, including smooth asymmetries and intricate correlations, making them better suited for capturing the nuanced structure in stellar parameters.
   
   For an illustration of a basic normalizing flow model, see Figure 2 from \citet{Weng_2018}\footnote{https://lilianweng.github.io/posts/2018-10-13-flow-models/}. For a more in-depth explanation of normalizing flows, see \citet{normflowOverview}, \citet{Weng_2018}, \citet{hon2024flowbasedgenerativeemulationgrids}, \citet{dinh2017density}, and \citet{Ting2022ApJ}.

    \subsection{Implementation of the StarFlow Model}\label{sec:ourmodel}

     We implemented a normalizing flow model using the RealNVP \citep[Real-valued Non-Volume Preserving,][]{dinh2017density} architecture. This architecture utilizes a stacked sequence of invertible bijective transformation functions (functions where each input is associated with a unique output). Each transformation function, or bijection  \(f:x \rightarrow y\), is implemented as an affine coupling layer. In each affine coupling layer, the input dimensions are split into two parts: the first \(d\) dimensions remain unchanged, while the remaining dimensions undergo scale and shift transformations, also known as affine transformations:
     \begin{displaymath}
         y_{1:d} = x_{1:d}
     \end{displaymath}
     \begin{equation}
         y_{d+1:D} = x_{d+1:D} \odot \exp (s(x_{1:d})) + t(x_{1:d}).
     \end{equation}
     Here, \(s\) and \(t\) are the scale and translation functions, respectively.  These functions can be arbitrarily complex and are typically modeled using a standard deep neural network. The RealNVP architecture is particularly well suited for modeling complex distributions while maintaining computational tractability.

     For our use case, we built a model that begins with a 6-dimensional Gaussian as the base distribution (\(p_0(z_0)\)). Each dimension corresponds to one of the six input parameters for which we aim to model the joint probability distribution function (PDF): effective temperature (\(T_{eff}\)), surface gravity (\(\log g\)), iron abundance ([Fe/H]), carbon abundance ([C/Fe]), nitrogen abundance ([N/Fe]), and stellar age or mass. 
     
     The model then applies a series of six flows, each consisting of an Affine Coupling Layer followed by a Permutation Layer. In our implementation, each Affine Coupling Layer is driven by a deep neural network with a 3x16x16x2 architecture. The first three neurons are responsible for processing the last three dimensions of the data (\(x_{d+1:D}\)), while the final two neurons output the scale and translation parameters. The Permutation Layer reorders the dimensions of the data, ensuring that all dimensions are transformed across multiple layers, thereby enhancing the model’s flexibility and expressiveness.

     In training machine learning models, a loss function is used as a measurement to guide training progress. The goal is to minimize this loss function. For our normalizing flows this translates to minimizing the negative log-likelihood. In practice, the loss function is calculated on batches of data points, so for the loss function we specifically use the forward Kullback-Leibler Divergence (KL divergence) between the empirical distribution of the data and the distribution modeled by the normalizing flow. The KL divergence quantifies how one probability distribution diverges from another, expected distribution. For a given set of observed stellar parameters \(x\), the forward KL divergence is defined as:

    \[\text{KLD}(p_{\text{data}} \parallel p_{\text{model}}) = \int p_{\text{data}}(x) \log \frac{p_{\text{model}}(x)}{p_{\text{data}}(x)} \, dx
    \]
     
     To implement this model in Python, we utilize the \texttt{normflows} package by \citet{Stimper2023} which is built upon the PyTorch framework \citep{paszke2017automatic}. Further discussion of the training process is in Section \ref{training}

\section{Data} \label{sec:data}

For this work we used the stellar parameters and abundances, and their uncertainties from ASPCAP, as well as stellar masses and ages from APOKASC 3 \citep{apokasc3}, and APO-K2 (\citealt{APOK2Age_2024AJ, APOK2Main_2024AJ}). In this section, we provide an overview of these data sources and include some details on the data reduction and calibration processes employed.

    \subsection{MWM \& APOGEE}
    Our spectroscopic data come from the SDSS-V MWM DR19 (Kollmeier et al. in preparation). Specifically, we utilize the stellar parameters (\(T_{\rm{eff}}\) \& log \textit{g}) and abundances ([Fe/H], [C/Fe], \& [N/Fe]) from APOGEE derived by the ASPCAP pipeline (\citealp{ASPCAP2016}, M{\'e}sz{\'a}ros et al. in preparation). APOGEE is a high-resolution (\(R \approx 22,500\)) H-band spectrograph that forms part of the MWM panoptic survey of SDSS-V. 

    Two separate subsets of APOGEE data are utilized for distinct purposes. Firstly, a subset of MWM stars with available mass and age information is employed to train our models on the full 6D distribution of \(T_{\rm{eff}}\), (\(\log g\)), [Fe/H], [C/Fe], [N/Fe], and age/mass. A piece of this same subset is also employed to verify the model's performance, as discussed in more detail in Section \ref{training}. Secondly, the much larger full set of MWM stars is used with the trained model to estimate masses and ages. This second set is not used during the training process and will be discussed more in Section \ref{results}.

    For both tasks, we exclude any stars with the \textit{BADSTAR} flag set. Additionally, we utilize the reported uncertainties for each of the five ASPCAP parameters, both in training the model and in recovering the full age posterior for MWM stars.

    \subsection{APOKASC \& APO-K2}
    For the asteroseismic data, we utilize data from the APOGEE-Kepler Asteroseismology Science Consortium \citep[APOKASC][]{apokasc3} and the APO-K2 catalog (\citealt{APOK2Age_2024AJ, APOK2Main_2024AJ}). Both data sets capture stars observed by both the MWM survey and the Kepler telescope. The main difference between these catalogs is in their observational histories: Kepler's initial 4-year single-field observation versus the subsequent 90-day K2 periods across multiple fields along the ecliptic. This variation in the duration affects the precision of the asteroseismic parameters and the range of star sizes that can be analyzed. APOKASC provides precise parameters for stars with relatively large radii, down to (\(\log g\)) = 1, but is restricted to one field. Conversely, APO-K2 observes multiple ecliptic fields, but has few asteroseismic parameters for stars above the red clump.

    We utilize the best age estimate columns from both catalogs along with their reported uncertainties. Asteroseismology is able to determine stellar ages by using asteroseismic scaling relations to determine the stellar mass, which is then used along with stellar isochrones to derive the age. APOKASC \citep{apokasc3} used the Garstec models \citep{garstec2008}, so our age estimations are also tied to these models. Users of our work may wish to have ages that are not tied to the predetermined isochrones from our asteroseismic training data, so we also utilize the best mass estimate columns in order to train and run a separate model that estimates stellar mass. This will allow users to utilize their own isochrones for mass to age conversion.
    
    The expanded asteroseismic datasets from APOKASC 3 and APO-K2 now include stars with metallicities as low as \([Fe/H] = -1.2\), allowing our model to work for stars with \(-1.2 < [Fe/H] < -0.5\) undergoing extra-mixing processes. We exclude stars with \([Fe/H] < -1.2\) due to the already sparse sampling of asteroseismic data for stars of this metallicity. Additionally, the age and mass datasets are trimmed to eliminate stars with relative age uncertainties exceeding 25\%.

    


    \subsection{Clusters}
    For our age validation stage we include comparisons to cluster ages. We use the cluster membership information from the Open Cluster Chemical Abundances and Mapping catalog \citep[OCCAM;][]{Myers_2022}, and obtained the cluster ages from \citet{clusterAges_Cantat-Gaudin}. The cluster ages have an associated uncertainty of \(\pm 0.15\) log(Age [Gyr]).
    \subsection{Training set}

\begin{figure*}[ht!]
\gridline{
  \fig{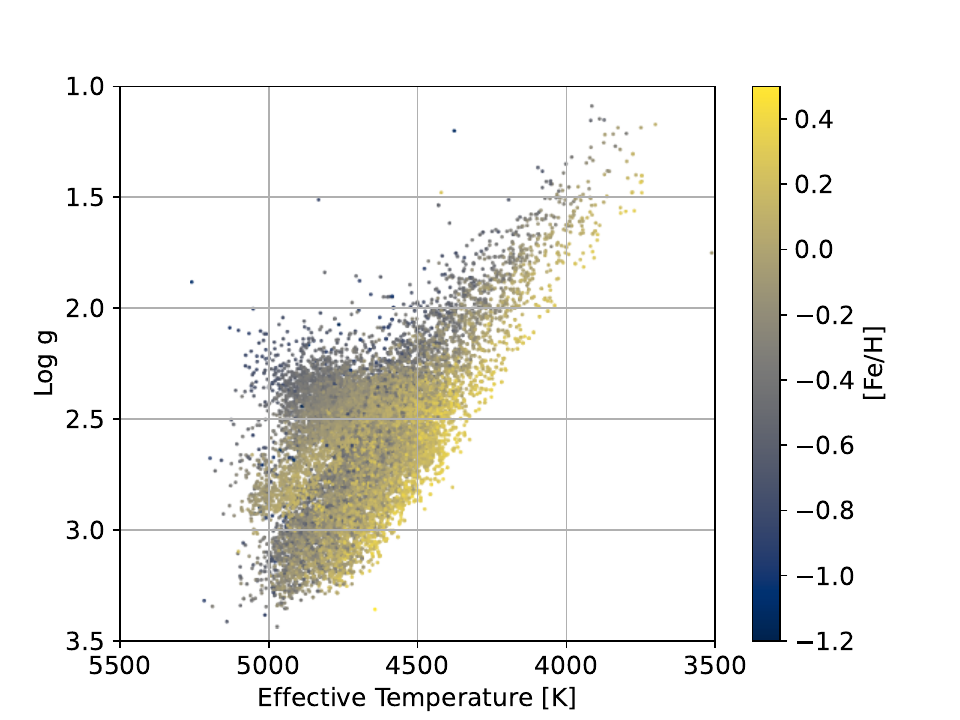}{0.45\textwidth}{}
  \fig{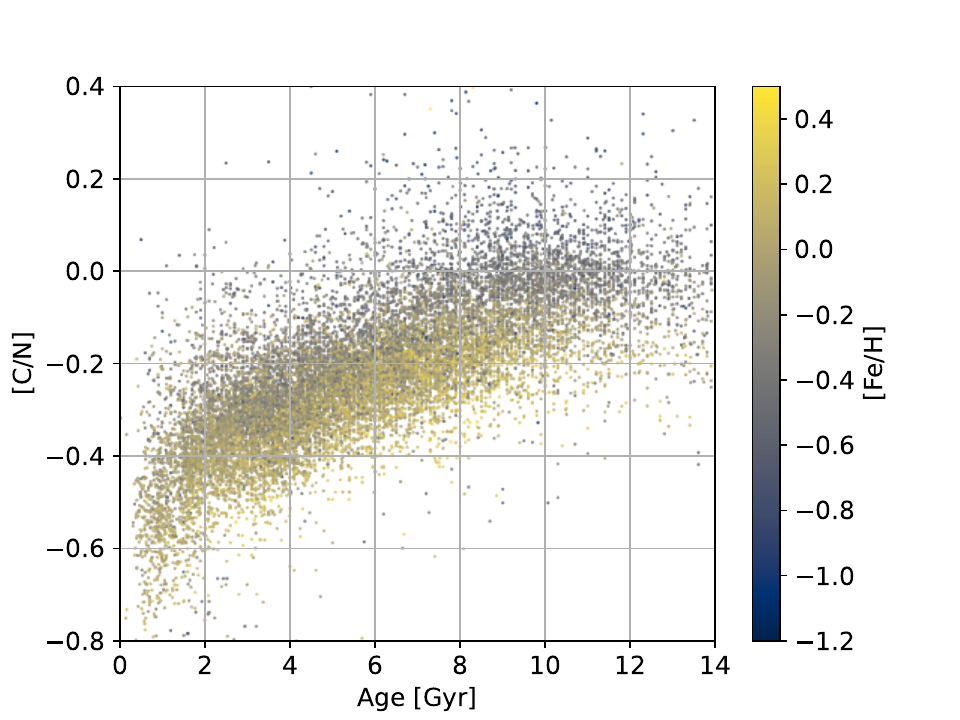}{0.45\textwidth}{}
}
\caption{Parameter space coverage of the training data. (\textbf{Left}) Kiel diagram of the training set color coded by [Fe/H]. (\textbf{Right}) [C/N] - Age relation of the training set color coded by [Fe/H].}
\label{fig:ParamSpace}
\end{figure*}
    
    Figure \ref{fig:ParamSpace} shows the parameter space coverage of the full combined data set. In general, the dataset has poor coverage in the upper RGB, with no stars at \(\log g < 1\). The age range of the dataset is from \(<1\)Gyr to \(14\) Gyr. It is important to note that asteroseismology can report stellar ages over a Hubble time due to the uncertainties in the measured asteroseismic parameters. Another important detail of the training set is the [C/N] - age relation plateaus at ages \(>8\) Gyr, as shown in the right-hand plot of Figure \ref{fig:ParamSpace}. This impacts the age estimation performance for older stars by increasing the size and skew of the estimated age posteriors, and is explored more in \citet{Roberts_2024MNRASR}. The final data set contains 15,641 stars.

\section{Training \& Application}\label{trainingAPP}
In the following subsections, we describe the process of training our normalizing flow model in (Section \ref{training}) and the procedures for sampling and utilizing the trained models in (Sections \ref{age} \& \ref{2dCPf}).

    \subsection{Model Training and Optimization}\label{training}

    The training process starts by augmenting the original training dataset to incorporate the labeled uncertainties associated with each parameter. This is achieved by sampling each star's parameters with Gaussian distributions centered on the reported value and with a standard deviation corresponding to their uncertainty. It should be noted that our model assumes uncorrelated errors between the parameters which is likely untrue. We assumed this due to there being no correlation matrix provided in the DR19 parameter file. Each star's distribution is sampled 500 times, which yields an total augmented dataset size of 7,820,500 samples. As a result, while the model may predict with broader error margins, these predictions are more robust and reflective of the true data distribution encountered in astronomical observations.
    
    The augmented dataset then undergoes standardization, where each parameter's mean is subtracted and then divided by its standard deviation. This standardized data is subsequently split into training, validation, and test sets with ratios of 0.7, 0.25, and 0.05, respectively. A blank model, as described in Section \ref{sec:ourmodel}, is initialized with PyTorch's default uniform initialization  (i.e., weights drawn from \([-1/ \sqrt{n},1/\sqrt{n}]\)), and the augmented and standardized training dataset is organized into batches of \(2^{16}\) (65,536) samples, which are used directly in each training epoch. No weight adjustments are applied to any of the training stars. During each epoch, the model computes the forward KL loss for the current training batch as well as for the validation set. The Adam optimizer is then used to update the model parameters based on the training batch loss, and the loss from the validation set is recorded. This validation loss serves as a metric to monitor training progress using data that are not included in the training set. Training is halted once the validation loss ceases to decrease for 100 consecutive epochs. Our model took \(\approx80,000\) epochs before ceasing.

    With the model now trained, both the standardization parameters and the test set are saved. The standardization parameters are crucial for applying the model to new data, ensuring consistency in data treatment. The test set, reserved for later use, will be utilized to verify the model's performance, as detailed in Section \ref{varification}.

        \subsection{Deriving Age Posteriors from Normalizing Flows}\label{age}

        The normalizing flow model calculates a log probability for any given sample point. To derive the probability distribution for stellar age, we employ the model to compute the conditional probability of age for a single star, given its stellar parameters and abundance information. This process involves creating an array of sample points that along one dimension is a linear grid of ages across a defined range of 0 - 14 Gyr, while the other dimension contains stellar property samples for a single star drawn from a Gaussian centered on the star's reported stellar properties and a width given by the reported uncertainties on those properties. Each star has its parameter distributions sampled 50 times. The model then computes the log probability for each age value in this array, then subsequently converts it to a probability likelihood. The probability likelihood is for the entire trained parameter space, so it is normalized to produce the age likelihoods used in age estimation. The age likelihoods across all the stellar property samples are then averaged to produce the final age probability distribution function for the star.
        
        To further refine these age distributions, we apply a simple flat age prior with zero probability at age\(>14\)Gyr and equal probability below, constraining the ages to not exceed a Hubble time. The final distribution is our age posterior, from which we can derive the maximum likelihood estimate of each star's age and determine the associated uncertainties.

\begin{figure*}[ht!]
        \plotone{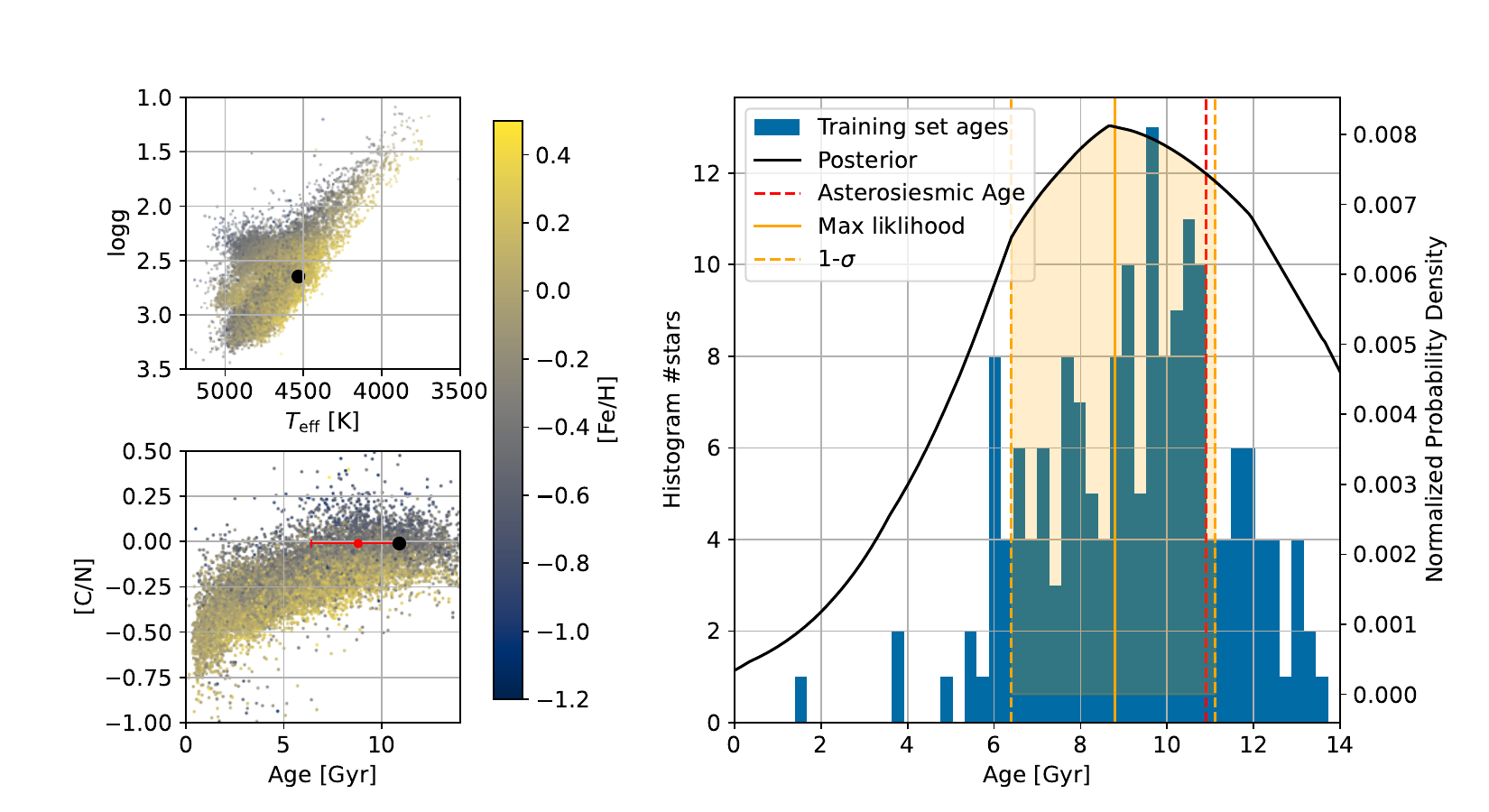}
        \caption{(\textbf{Left}) The larger black point marks the location of an example test-set star overlaid on the full training set (color-coded by [Fe/H]) shown in a Kiel diagram (\textbf{Upper Left}) and in the [C/N]–age relation (\textbf{Lower Left}). In the [C/N]–age plot, a red error bar indicates the model's age estimate and uncertainty for the star. (\textbf{Right}) Age posterior for the example star. The black curve represents the full posterior computed with the input uncertainties, with the solid orange line indicating the maximum likelihood and dashed lines showing \(1\sigma\) error bars. The red dashed line indicates the asteroseismic age, while the blue histogram displays the age distribution of training-set stars that are nearby in input parameter space. Additional example stars from the test set are shown in Appendix \ref{sec:examples}}
        \label{fig:AgePosterior}
        \end{figure*}

        Figure \ref{fig:AgePosterior} shows a representative test-set star with asteroseismic age near the older end of the distribution. Additional examples for young and intermediate-age stars are shown in Appendix \ref{sec:examples} for comparison. The relation between the orange curve and the blue histogram shows how the recovered age posterior approximates the underlying distribution of ages in the training set. The histogram does not account for input uncertainties in the data, however the recovered posterior shown in orange does. This example happens to just barely succeed in recovering the age to \(1\sigma\) and demonstrates how the posterior from normalizing flows can have asymmetric uncertainties.

        Using the recovered age posterior, we assign a maximum-likelihood age for each star along with \(1\sigma\) age uncertainties. The age uncertainties are determined by identifying the age values that lie symmetrically around the posterior peak and encompass 68\% of the cumulative probability. Specifically, we first locate the peak of the posterior distribution, representing the most probable age estimate. We then compute the cumulative distribution function (CDF) of the posterior and determine the lower and upper bounds where the CDF decreases and increases by half of the desired percentile relative to the peak's CDF value. These bounds effectively capture the central 68\% of the probability distribution, accounting for both symmetric and asymmetric uncertainties.

        \subsection{Isochrone-Agnostic Age Estimation}\label{mass}

        We use the same approach as the model trained on stellar ages to produce mass posteriors for each star, with the one significant distinction being that the mass model is trained on \(\log(M_\odot)\). Our motivation for this is to allow users of the catalog to utilize alternative isochrones than those used by APOKASC to determine stellar ages \citep{garstec2008}. 
        

        


        \subsection{Model Predictions Beyond Stellar Ages}\label{2dCPf}
       The normalizing flow model allows simultaneous sampling of multiple parameters, which is particularly advantageous for evaluating the [C/Fe] and [N/Fe] ratios in stars. For stars with known asteroseismic masses in the test set, we generate two-dimensional probability distributions of [C/Fe] and [N/Fe]. This process parallels the mass posterior recovery method, except that we keep the mass fixed at the asteroseismic value along with the stellar parameters, allowing the [C/Fe] and [N/Fe] abundances to vary freely. So each star's [C/Fe] \& [N/Fe] probability distribution uses the \(T_{\rm{eff}}\), (\(\log g\)), [Fe/H], and asteroseismic age as inputs. The resulting 2D sample grid, when analyzed by the model, produces a 2D probability distribution with axes for [C/Fe] and [N/Fe], illustrating the model’s predictions for these ratios in a given star.

        \begin{figure}[ht!]
        \gridline{
          \fig{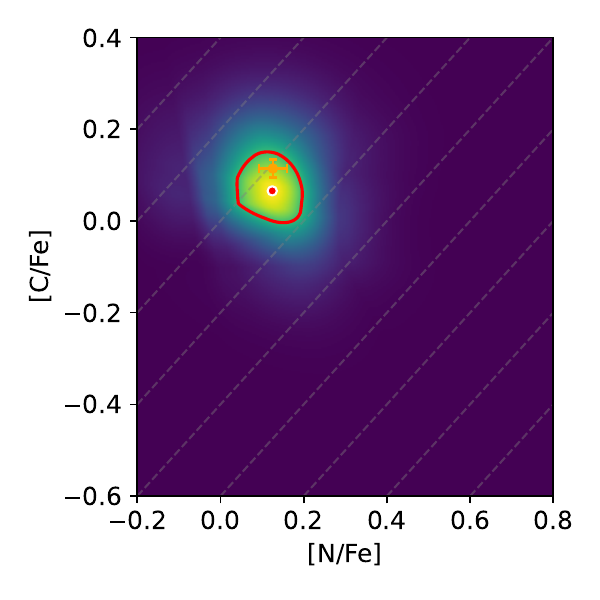}{0.75\columnwidth}{}}
      \gridline{
          \fig{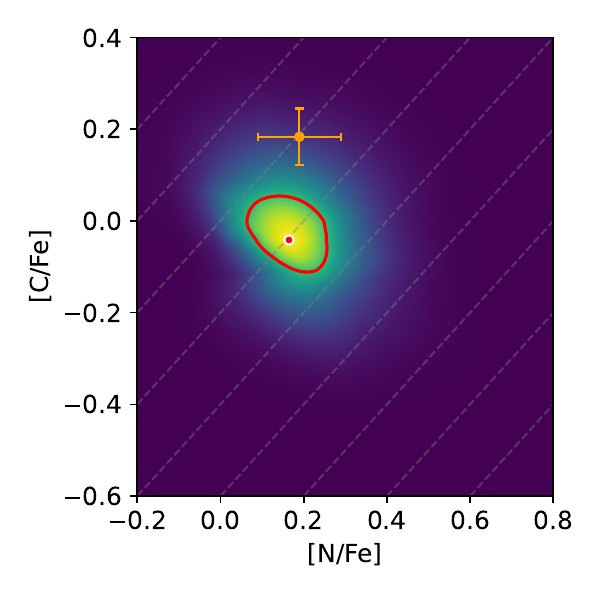}{0.75\columnwidth}{}
        }
        \caption{(\textbf{Upper}) 2D probability distribution of [C/Fe] \& [N/Fe] for the star depicted in Fig \ref{fig:AgePosterior}. The red dot marks the maximum likelihood, and the red line indicates the \(1\sigma\) boundary. The orange point and error bar shows the star's measured [C/Fe] \& [N/Fe] and associated uncertainties. The diagonals show lines of constant [C/N]. (\textbf{Lower}) Displays a similar plot for a different star with a higher [C/Fe] than the model predicts. Both plot exhibit slight banding features, these are most likely artifacts of the normalizing flow model over fitting to local noise or small-scale features.}
        \label{fig:CNcond}
        \end{figure}

        This analysis aids in evaluating the [C/Fe] \& [N/Fe] abundances and in detecting atypical abundance patterns. Figure \ref{fig:CNcond} illustrates two examples of C \& N probability distributions. The left plot shows the C \& N distribution for the star in Figure \ref{fig:AgePosterior}, while the right plot illustrates the distribution for another star with an unexpectedly high reported [C/Fe]. The star shown in the right plot is both at a higher [C/N] than expected indicating that it is younger than the [C/N] - age relation would predict, and it is carbon enriched compared to similar stars of its age, evolutionary type, and [Fe/H]. A more comprehensive analysis of the entire test set using this methodology is detailed in Section \ref{CandN}. 
        
        For stars lacking a known mass, the analysis could be expanded to a three-dimensional sample grid, varying C, N, and mass. We then combine the C \& N data to explore the model's predictions regarding the [C/N]-mass relationship across parameter space. This approach would help identify stars with atypical C and N abundances at any prospective mass, providing a less precise yet useful method for identifying stars whose chemical compositions may be unreliable indicators of age. 

        \subsection{Training Space Density}\label{sec:density}
            
            An additional feature of normalizing flows mentioned in Section \ref{sec:motivationMethod} is that the model also retains information about the distribution of the training data in parameter space. Essentially, when we recover the unnormalized age posterior, we are also sampling the underlying training space density at each point. This density value is significantly larger than the original training sample size due to the stars being sampled multiple times per training epoch and the large number of epochs.
            

            \begin{figure}[ht!]
                \gridline{
                  \fig{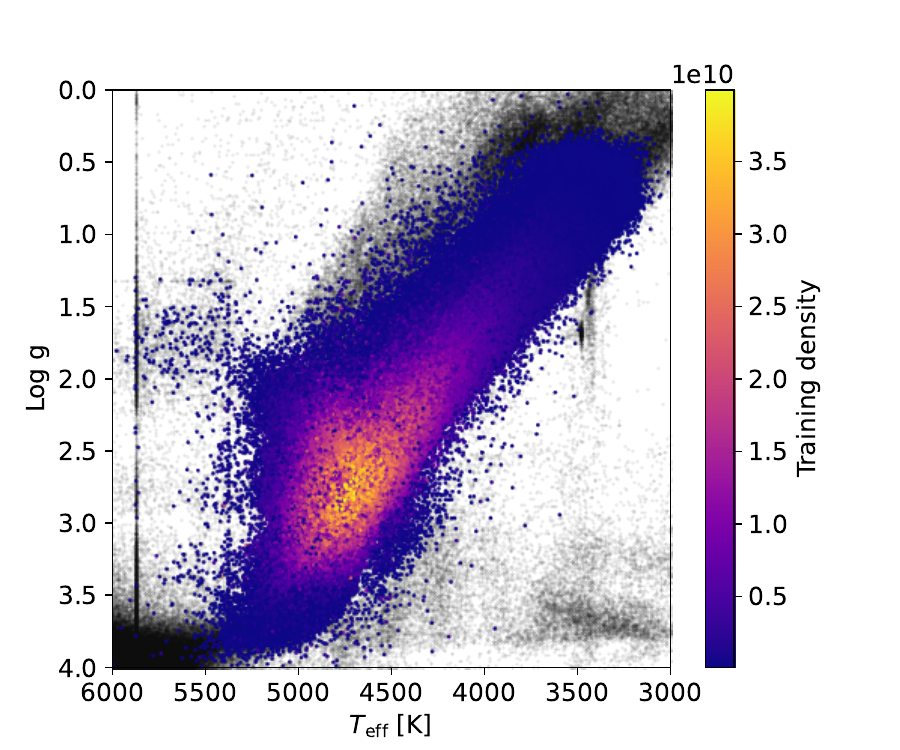}{0.9\columnwidth}{}}
              \gridline{
                  \fig{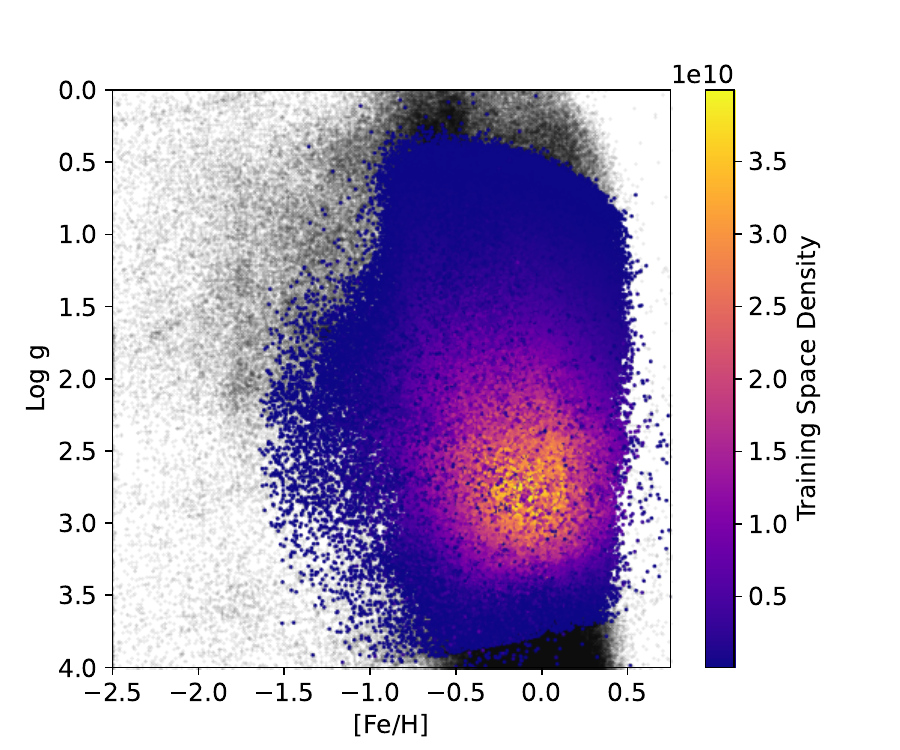}{0.9\columnwidth}{}
                }
                \caption{Parameter space coverage of stars with a training space density above \(3\times10^9\) (color-coded by density) compared to the full DR19 dataset (black points). These are scatter plots with each stars training density used for color coding. Regions with higher densities are better sampled in the training data, yielding more reliable age estimates.  \textbf{(Upper)} Kiel diagram \textbf{(Lower)} [Fe/H] versus surface gravity (\(\log g\)). The sharp banded features seen in the black points are artifacts from the ASPCAP grid edges. This Figure is repeated in Appendix \ref{sec:densityfigure} with different training space density thresholds.}
                \label{fig:Density}
            \end{figure}

            We utilize this value as a measure of how well-sampled the parameter space around a given star’s parameters was during training, and we use it as a cutoff to limit age estimates to stars that occupy well-covered regions of parameter space. Figure \ref{fig:Density} shows a Kiel diagram of the stars in our catalog, color-coded by their training space density. The right-hand panel shows the same sample but restricted to stars with a training space density greater than \(3\times10^9\), closely reflecting the parameter space coverage of the original training set. Choosing a smaller cutoff allows the model to extrapolate into less well-sampled regions, while a higher cutoff restricts the sample to stars within the most densely populated parts of parameter space. 
            In subsequent sections, we refer to this value as the training space density, and we adopt a cutoff of greater than \(3\times10^9\) for the age validation. This choice demonstrates confidence in the model’s ability to extrapolate slightly beyond the asteroseismic core of the training data. We feel that this adopted value is a good starting point that is not too conservative, resulting in a small number of stars in the output, while also constraining the model from extrapolating too far. When applied to the training set, this value results in 95\% of the training data being retained. Section \ref{ages} and Figure \ref{fig:agehistall} further discuss the effects of this decision and alterations in star counts for the whole DR19 sample. Appendix \ref{sec:thresholdChange} shows how varying the training space density used as a threshold changes the various plots and maps shown in this work.

        \section{Validation}\label{varification}
        Age estimations of field stars are difficult due to dependencies on stellar models and sizable age uncertainties. This makes systematic age biases unavoidable and challenging to quantify. The best method to validate our ages is to make a series of comparisons of our model ages against multiple literature sources that use different methods to determine ages. Using these comparisons, we also provide validation of the uncertainties produced by the model.
    
            \subsection{Validation Against Asteroseismic ages}\label{AsteroseismologyCompare}
\begin{figure}[ht!]
                \centering
                \includegraphics[width=0.95\linewidth]{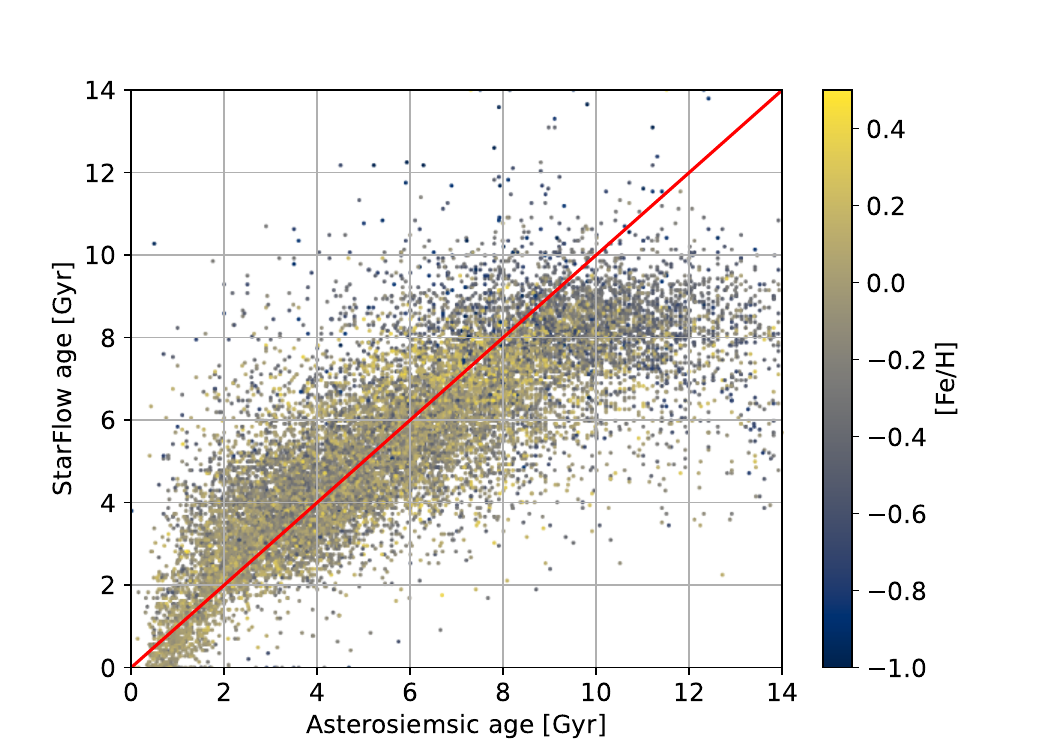}
                \caption{Scatter plot comparing model-derived maximum likelihood ages to asteroseismic ages for the test set. Points are color-coded by iron abundance ([Fe/H]). The red line indicates the 1-1 relation where both ages agree.}
                \label{fig:AgeVsAge1}
            \end{figure}

            Figure \ref{fig:AgeVsAge1} shows the maximum likelihood ages for the test set as determined by our model compared to the asteroseismic ages. The figure demonstrates that for ages \(< 8\) Gyr the model returns accurate ages. However, at ages \(> 8\) Gyr, the model's maximum likelihood begins to diverge from the asteroseismic estimates. This result is not unexpected, as it has been observed and documented in multiple spectroscopic age studies (\citealt{AstroNN, MackerethAges, AndersAges, DistMass}). The plateau at 8 Gyr is also expected from [C/N] ages, as shown in \cite{Roberts_2024MNRASR}, due to mixing processes being weaker in lower mass (older) stars. Another potential contributor to the observed behavior is contamination of the training set by stars that have experienced mergers or mass transfer. Such stars can exhibit atypical [C/N] ratios for their true age. However, these objects constitute a small fraction of the training set and are unlikely to significantly shift the maximum likelihood estimates produced by the model.
        
\begin{figure}
            \centering
            \includegraphics[width=0.95\linewidth]{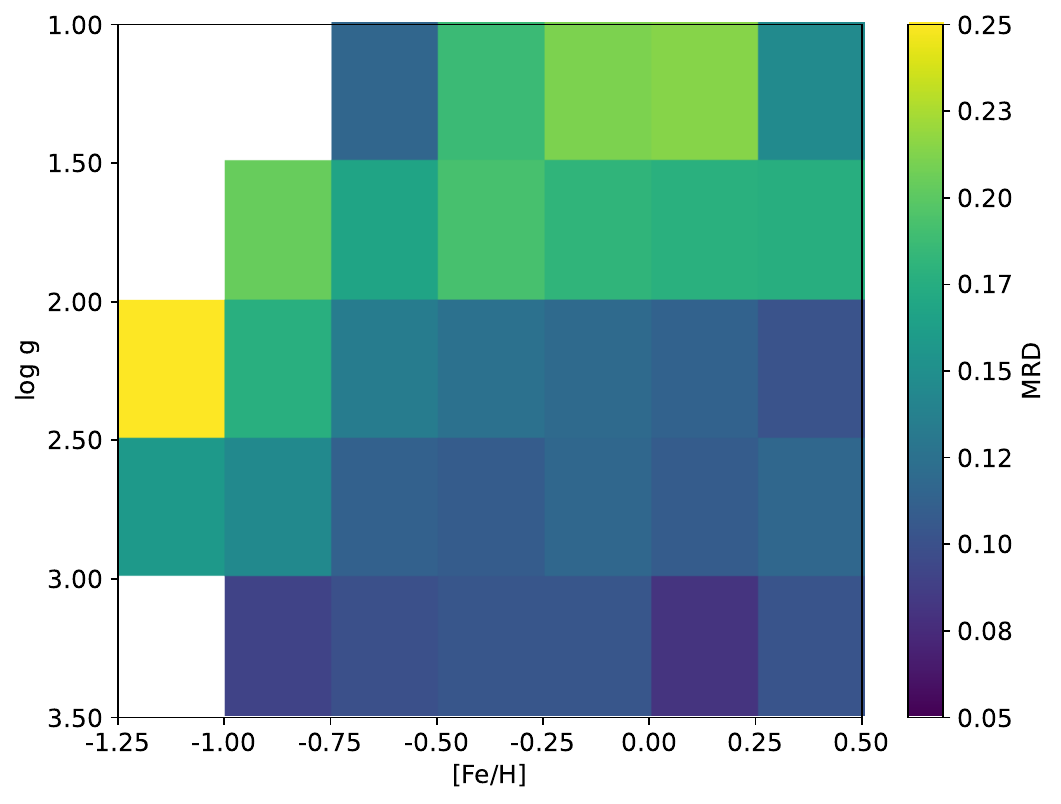}
            \caption{Median Relative Deviation (MRD) of model-derived ages compared to asteroseismic ages in the test set, binned by \(\log g\) and [Fe/H]. A high MRD indicates that there is a larger scatter in the residual of the StarFlow ages vs the asteroseismic test set ages. In general we see that the MRD is lower for stars at higher \(\log g\) and [Fe/H].}
            \label{fig:MRD}
        \end{figure}

           To assess the scatter between the model's maximum likelihood ages and the asteroseismic ages, we present the Median Relative Deviation (MRD) in Figure \ref{fig:MRD}. The plots display the MRD binned across [Fe/H] and (\(\log g\)) to examine how the scatter varies with these parameters. For [Fe/H], the MRD remains relatively constant, with only a slight increase in the MRD for metal-poor stars.  In contrast, variations in (\(\log g\)) show a more significant change: lower RGB stars exhibit a scatter of approximately  $\approx 11 \%$ while upper RGB show a scatter of $\approx23\%$. 
    

            Both of these results are not unexpected. Metal-poor stars are less represented in the training set, and there are potential extra-mixing effects at [Fe/H] \(< -0.5\) that could introduce more scatter in the maximum likelihood ages \citep{ShetroneMixing}. Additionally, upper RGB stars are also poorly represented in the training set, and asteroseismic mass determination for these stars is much more difficult due to larger uncertainties.

            \subsection{Evaluating Model-Derived Uncertainties}
    
            Typically, spectroscopic age catalogs use the median relative deviation or other statistics that compare the scatter of predicted ages to asteroseismic ages as the basis for their reported uncertainties (\citealt{AstroNN, DistMass, MackerethAges}). Our process for determining the complete age posterior, which accounts for all sources of uncertainty including the model uncertainty, is described in Sections \ref{training} and \ref{age}. The primary motivation for utilizing a normalizing flows methodology was to obtain individual uncertainty measurements for each star. This approach leverages the uncertainties of all measured parameters during both the training process and the age estimation for each star.


        \begin{figure}[ht!]
            \gridline{
              \fig{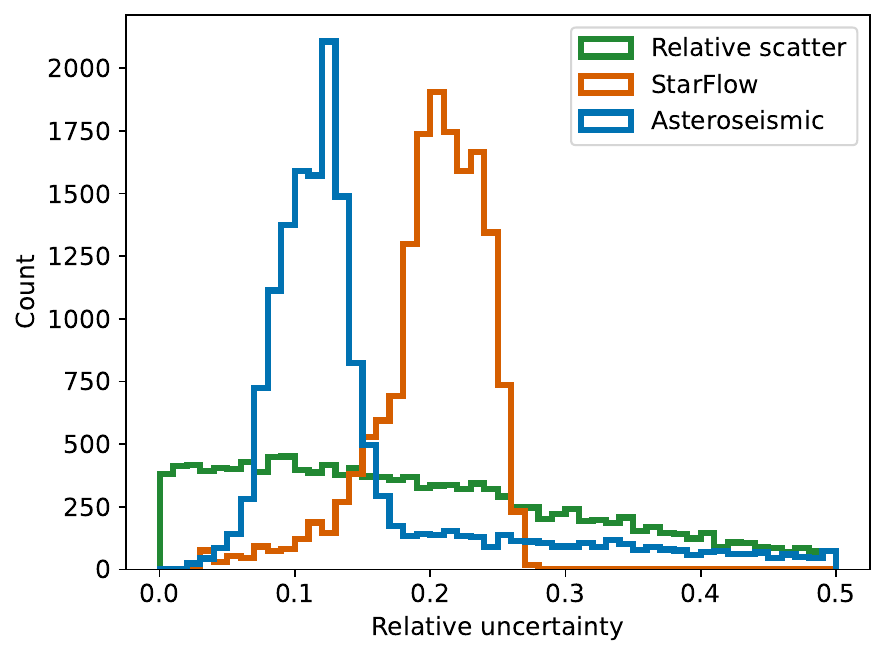}{0.9\columnwidth}{}}
              \gridline{
              \fig{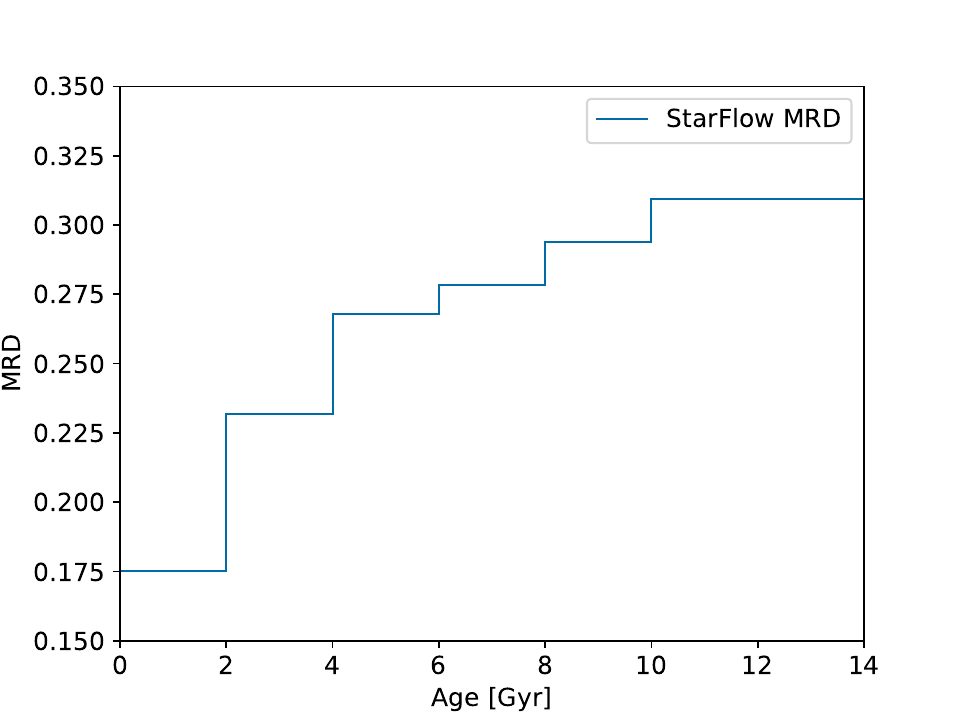}{0.9\columnwidth}{}
            }
        \caption{(\textbf{Upper}) Histogram of the relative $ 1\sigma$ uncertainties for  asteroseismic labels (blue), the averaged upper and lower relative $1\sigma$ uncertainties derived from the StarFlow model (orange), and the relative differences between the asteroseismic age and the StarFlow maximum likelihood ages (green). (\textbf{Lower}) Bar plot showing the median relative uncertainty recovered from the StarFlow age posteriors, binned by asteroseismic age.}
        \label{fig:UncertHist}
        \end{figure}

            Figure \ref{fig:UncertHist} (Left) shows a histogram of the uncertainties from the asteroseismic data, the derived uncertainties of our model from the posterior, and the relative difference between our model's estimated age and the asteroseismic age. The reported asteroseismic uncertainties are mostly around 12\% with some stars having uncertainties up to 50\% and higher. The relative differences in ages range from near 0\% to over 50\%. The relative differences shown here are typically used as the basis for derived uncertainties in most spectroscopic catalogs. Taking the median or mean of the relative difference results in a value between $\approx 10\%$ and 20\%. Using this value as the standard deviation results in only $\approx 50\%$ of the stars having asteroseismic ages within $1\sigma$ of the model estimations, whereas it should be closer to 68\% for $1\sigma$. Our model has a typical age uncertainty of $\approx 28\%$, though it varies for different stars. We determined that the $1\sigma$ uncertainties produced by our model capture $\approx 72\%$ of the asteroseimic ages. This implies that our model uncertainties are slightly overestimated. We also compute the Z-scores of our ages using $(\hat{y}-y) / \sqrt{\sigma_{y}^{2} + \sigma_{\hat{y}}^{2}}$ where $y$ is the ground truth age, $\hat{y}$ is the model age, and $\sigma_{y}$ and $\sigma_{\hat{y}}$ are their respective uncertainties. The resulting distribution should follow a standard normal distribution if our uncertainties are correct. However, we find that the resulting distribution is narrow with a standard deviation of 0.86, which also implies that our uncertainties are overestimated.
            
            Figure \ref{fig:UncertHist} (Right) shows the median age uncertainty from our model, binned by asteroseismic age. The uncertainty is $\approx 18 \%$ for younger stars and increases until ages $>8$ Gyr at which the uncertainty is $\approx 30\%$. This indicates that the model uncertainties are accounting for the breakdown of the [C/N] - age relation as discussed in Section \ref{AsteroseismologyCompare} and by \citet{Roberts_2024MNRASR}.
            
            \subsection{Cluster Ages as Validation Benchmarks}
\begin{figure*}[ht!]
                \centering
                \includegraphics[width=0.85\linewidth]{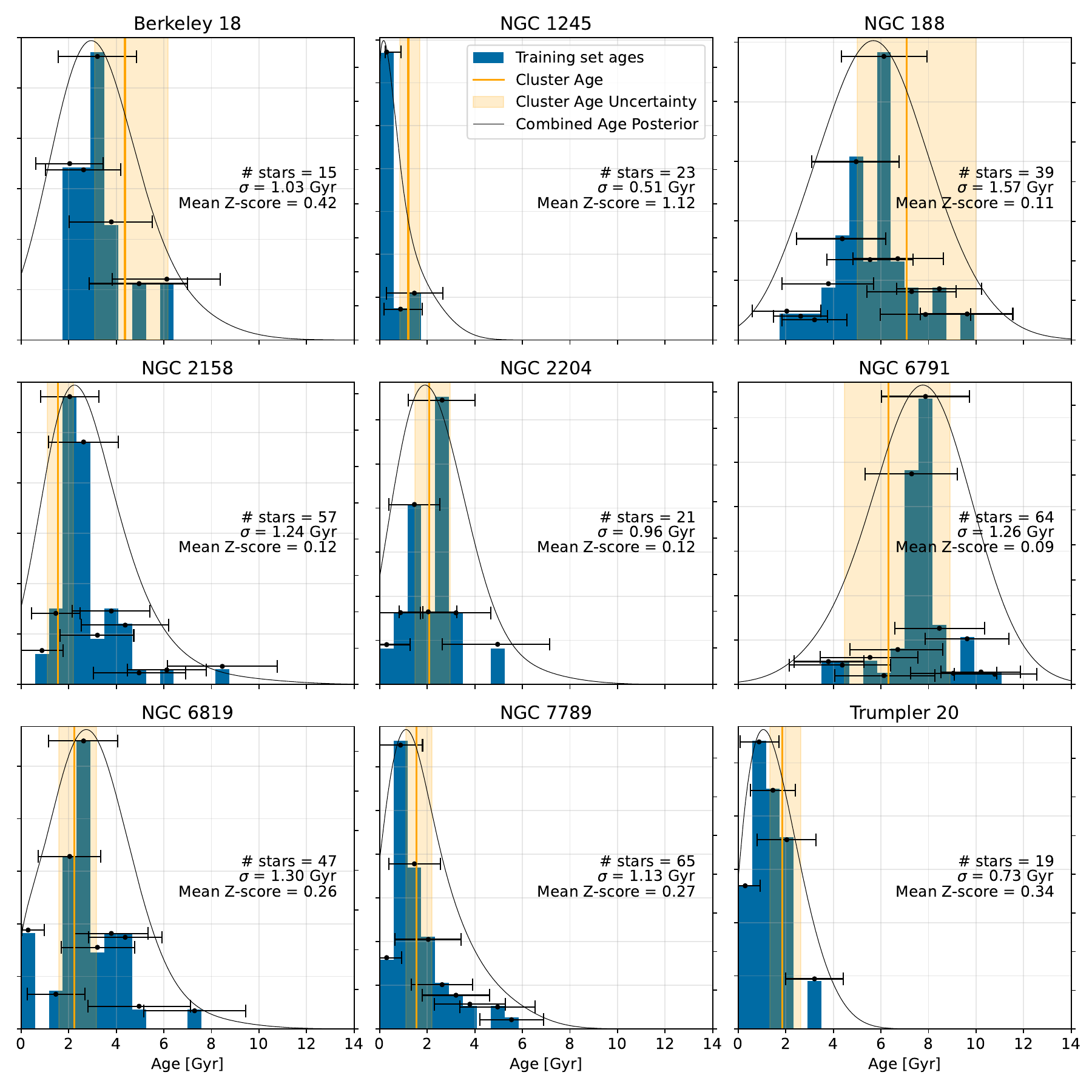}
                \caption{Set of histograms showing our age estimations for nine stellar clusters. Each bin is overlaid with error bar representing the mean positive and negative uncertainties for the stars within that bin. The orange line and shaded region indicate the cluster age and its uncertainties as reported by \citet{clusterAges_Cantat-Gaudin}. The black curve illustrates the combined age posteriors from our model for the stars in the cluster. Each histogram includes the number of stars matched to our catalog under the conditions of membership prob \(> 0.85\) \& Training density \(>3\times10^9\). Additionally, each plot shows the mean Z-score and its standard deviation \(\sigma\) for the estimated ages within the respective cluster.}
                \label{fig:ClusterHist}
            \end{figure*}

            Star clusters, which consist of stars that formed simultaneously, serve as excellent references for validating our age estimates. Figure \ref{fig:ClusterHist} illustrates the age distributions for several clusters from the Open Cluster Chemical Abundances and Mapping catalog \citep{Myers_2022}. The histograms show the maximum likelihood ages of stars from our catalog with a cluster membership probability of 85\% or higher. Each bar is overlaid with error bars representing the mean positive and negative uncertainties for the stars in each bin. The black curve represents the averaged age posterior for all stars in the cluster from our model, while the orange line and shaded region indicate the cluster age and its uncertainty from \citet{clusterAges_Cantat-Gaudin}.
                
            In general, our estimated ages match the cluster ages within \(1\sigma\). To quantify this agreement, we calculate a standardized difference  by subtracting the cluster’s age from each star’s age and dividing by the quadrature sum of their respective uncertainties. For most clusters, this standardized difference remains below 1, with NGC 1245 as the lone exception—a result not too surprising given that the [C/N]–Age relation breaks down for ages \(<1\) Gyr \citealp{Spoo_2022,Roberts_2024MNRASR}. In the model results, many NGC 1245 stars are placed at the lowest edge of the age grid, reflecting this breakdown. Across all clusters, the average standardized difference is 0.44, and within each cluster, the standard deviation of our derived ages ranges from 0.6 Gyr to 1.7 Gyr.



            A few clusters show some anomalous characteristics. Clusters like NGC 6819, 2158, 188, and 7789 exhibit slightly skewed or bi-modal distributions. These deviations could be due to a mix of stars with different chemistries, possible misclassification of cluster membership, or stellar interactions. In particular, NGC 6791 exhibits a large spread in estimated ages, with a large peak in age estimations being 2 Gyr older than the reported cluster age by \citet{clusterAges_Cantat-Gaudin}. However \citet{6791Clust} and \citet{6791_Brogaardage} place the age of NGC 6791 at \(\approx\) 8 Gyr which does match our model's age estimation. \citet{clusterAges_Cantat-Gaudin} likely underestimated the age due to their model not understanding NGC 6791's high metallicity \citep{6791_Brogaardage}.
            

            \subsection{Comparison with Existing Stellar Age Catalogs}
\begin{figure*}
                \centering
                \includegraphics[width=0.9\linewidth]{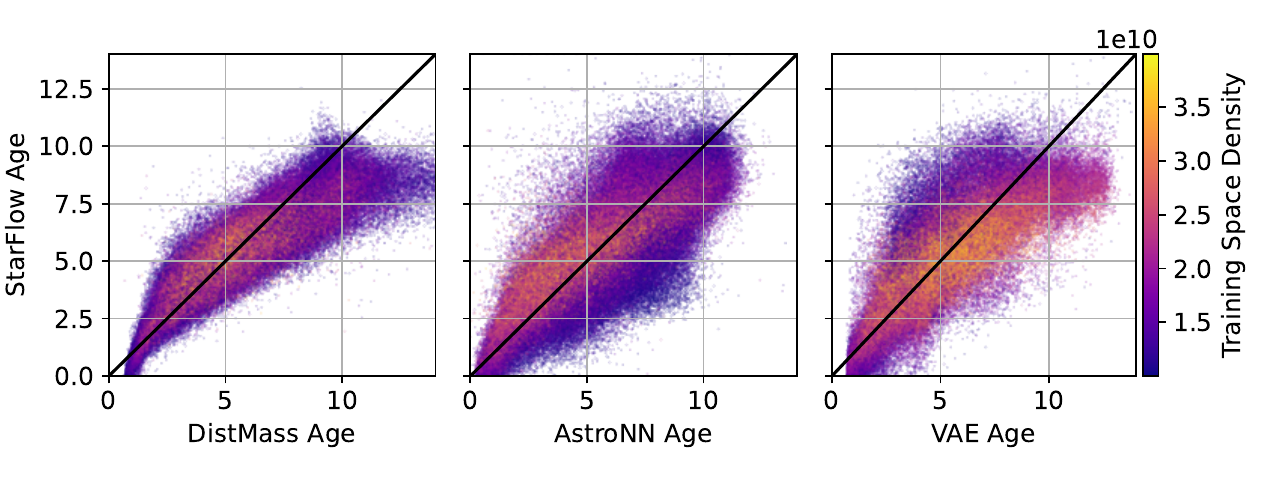}
                \caption{Comparisons of model-derived ages with other age catalogs.\textbf{(Left)} Scatter plot comparing the \textit{DistMass} ages from \citet{DistMass} to our model's maximum likelihood ages. \textbf{(Center)} Scatter plot comparing \citet{MackerethAges} ages to our model's ages. \textbf{(Right)} Scatter plot comparing \citet{Leung_2023MNRAS} ages to our model's ages. In each subfigure, the black line represents the one-to-one correspondence between the two age estimates. The color gradient indicates the training space density of our model for each star.}
                \label{fig:compare}
            \end{figure*}

            In \citet{DistMass}, we described the DistMass catalog, which derives stellar ages from spectroscopic parameters using a simple dense neural network. Figure \ref{fig:compare} shows different chemical age catalogs vs our StarFlow ages. The left panel shows a comparisons to the ages from DistMass. The center and right panels  present comparisons against \citet{MackerethAges} and \citet{Leung_2023MNRAS}, respectively.

            In general, our ages match the DistMass, AstroNN, and \citet{Leung_2023MNRAS} ages within \(1\sigma\). However, there are discrepancies in the comparisons against \textit{Distmass} \citep{DistMass}, and \citet{Leung_2023MNRAS} for the older stars at ages \(>8Gyr\) where the comparison appears to plateau. The discrepant stars in the \textit{DistMass} comparison have a \(T_{\rm{eff}}\) and \(\log g\) that place them in the red clump. This is not surprising as the more simple ML architecture of \textit{DistMass} likely did not learn the slightly different [C/N] - age relation that exists in the red clump. We do not believe the red clump poses a problem for the StarFlow model due there being no jumps or changes in the relative scatter of the StarFlow ages when compared to asteroseismology around the red clump. 

            The plateau feature in the age comparison to \citet{Leung_2023MNRAS} is more interesting, as it does not appear to have dependence to any stellar parameter other than the estimated age itself. It is similar to what we observed when comparing our ages to asteroseismic ages and is attributed to the [C/N]-age relation becoming less effective at older ages \citep{Roberts_2024MNRASR}. In contrast, \citet{Leung_2023MNRAS} reports that their ages do not exhibit this plateauing behavior, which may be due to their model better utilizing information from the spectra.

\section{Results}\label{results}

    \subsection{Stellar Age Estimates for SDSS-V DR19}\label{ages}
    By the end of SDSS-V, there will be millions of APOGEE spectra of unique stars across parameter space. It will be important to obtain accurate ages for a wide range of stars using different methods in order to use the full power of the final dataset. Our approach yields a comprehensive set of age estimates that incorporate confidence levels based on training parameter space coverage and associated age uncertainties for each star. As discussed in Section \ref{sec:density}, our estimations retain information about the training space density, which serves as a data selection criterion to control model extrapolation.
    
\begin{figure*}[ht!]
    \centering
    \includegraphics[width=1.1\linewidth]{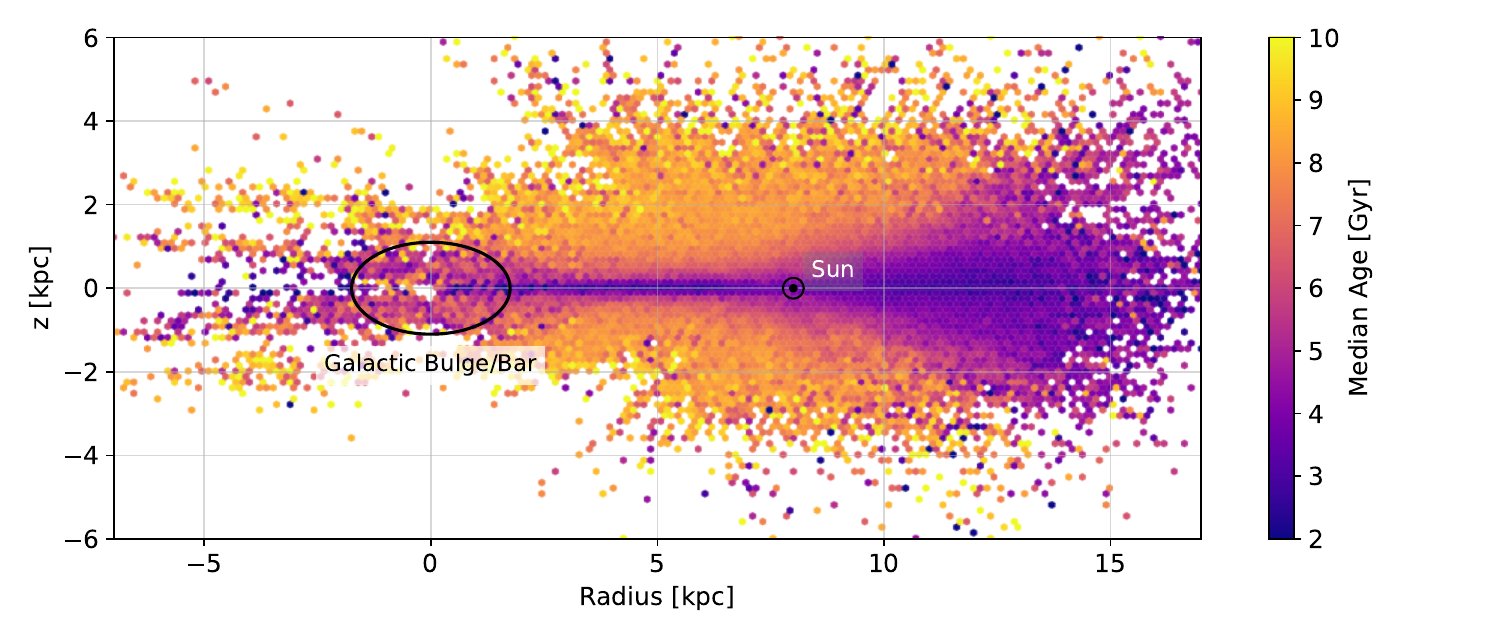}
    \caption{This annotated hexbin map illustrates the spatial distribution of the median of our stellar ages across the Milky Way, projected onto the R (Galactocentric radius) versus Z (height above the Galactic plane) plane. We used photogeometric distances from \citet{GaiaDistance}. Only stars with a training space density \textit{density} \(> 3\times10^9\) are included. Map is annotated with the Galactic center and the location of the sun. Regions with older median ages are depicted in warmer colors , while younger regions are shown in cooler colors. This visualization highlights the age gradients and structural features of the Galaxy, such as the distribution of younger stars in the thin disk and older stars in the thick disk. This figure is repeated in Appendix \ref{sec:agefigure} with different density thresholds to illustrate how changes to the threshold affect the age map.}
    \label{fig:agemap}
    \end{figure*}

    By applying our recommended density threshold of \textit{density} \(> 3\times10^9\), we have successfully estimated ages for 378,720 stars in SDSS DR19. This dataset is over two times larger than the 142,257 stars with age estimates from \citet{Leung_2023MNRAS}. Figure \ref{fig:agemap} presents an R-Z projection of the Galaxy using distances from \citet{GaiaDistance}, color-coded by the median ages of stars that meet the density criterion of \textit{training density} \(> 3\times10^9\). The training density value is discussed in Section \ref{sec:density}. This visualization highlights the spatial distribution of stellar ages across different regions of the Galaxy. It clearly shows the young thin disk, the flaring of the thin disk at higher \(R_{\rm{gal}}\), the old thick disk, the mixed ages of the Galactic core, and the well-documented age gradient in the Milky Way disk (\citealp{martigAgeGrad,Frankel_2019,Imig_2023})

\begin{figure}
    \gridline{\fig{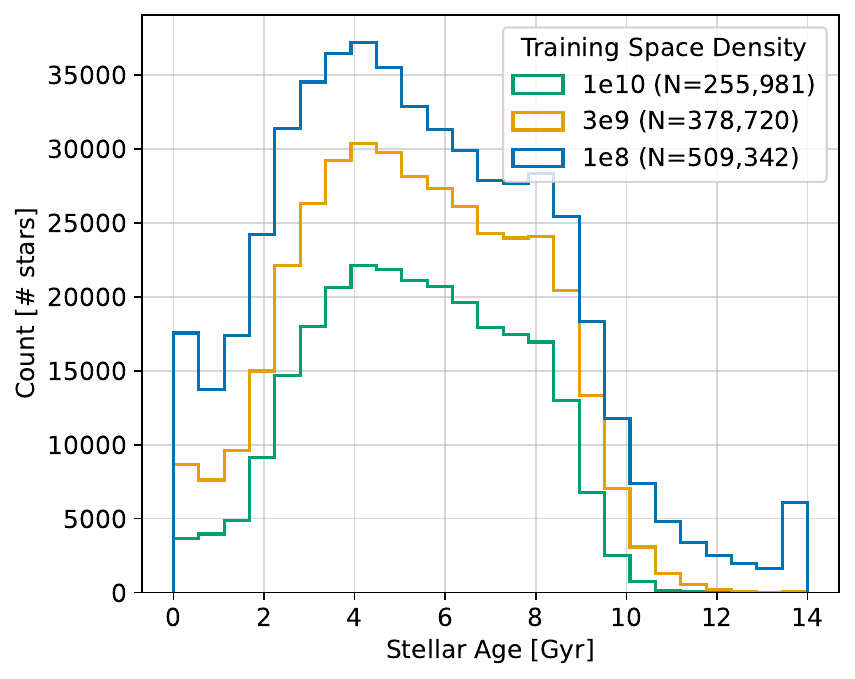}{0.9\columnwidth}{}}
    \gridline{
    \fig{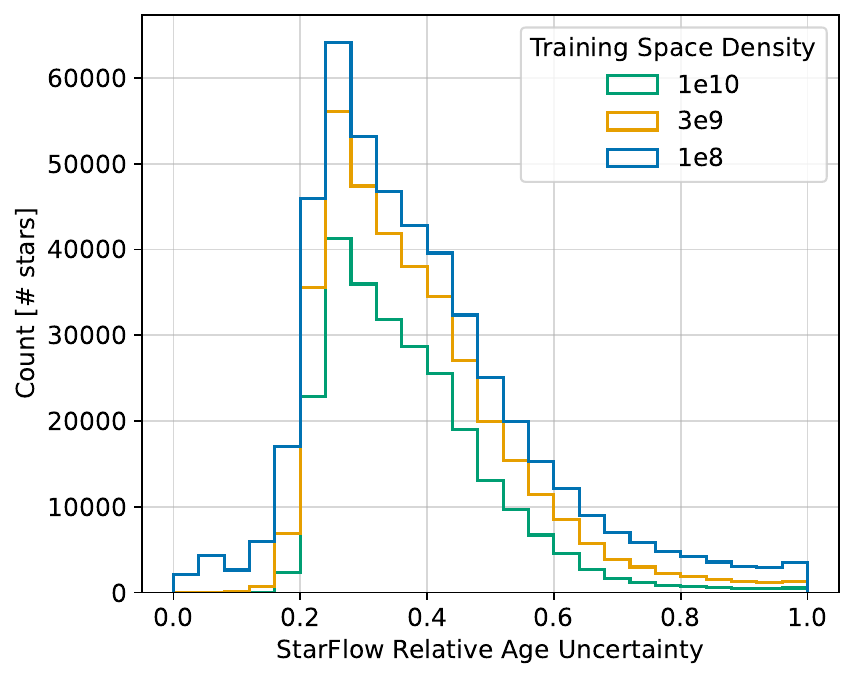}{0.9\columnwidth}{}}
    \caption{Histograms displaying the number of stars at each estimated age for three different training space density thresholds: \textit{density} \(> 10^8\), \textit{density} \(> 3\times10^9\), and \textit{density} \(> 10^9\). \textbf{(Upper)} The legend indicates the total number of stars included using each threshold. Lowering the density threshold increases the sample size but also leads to greater model extrapolation, as evidenced by the accumulation of age estimates at the grid edges (0 Gyr and 14 Gyr). \textbf{(Lower)} Histogram illustrating the distribution of reported age uncertainties for stars at the same density thresolds. These panels highlight the impact of density-based selection on the reliability and distribution of our age estimates within the MWM sample.}
    \label{fig:agehistall}
\end{figure}

    Figure \ref{fig:agehistall} displays histograms of stellar ages under three distinct density criteria:  \textit{density} \(> 10^8\), \textit{density} \(> 3\times10^9\), and \textit{density} \(> 10^{10}\). Lowering the density threshold allows for increased model extrapolation and Galactic coverage; however, this leads to less accurate age estimations. Specifically, when \textit{density} \(< 6\times10^8\), the model tends to cluster more ages at the grid edges, resulting in a significant number of stars being assigned ages of 0 Gyr and 14 Gyr.

    Additionally, Figure \ref{fig:agehistall} presents the distribution of age uncertainties across the same training density thresholds. Most stars exhibit reported age uncertainties ranging from approximately 20 to 40\%. Figure \ref{fig:agehistall} also displays a higher number of stars with greater uncertainties compared to the test set sample shown in Figure \ref{fig:UncertHist}, attributable to the larger and more varied parameter space covered by the entire MWM DR19 dataset. Notably, at higher density thresholds, there is a reduction in the number of stars with low reported age uncertainties  (\(< 0.1\)). However, these low uncertainty estimates predominantly correspond to stars located at the upper edge of the age grid, so they are likely artifacts of the model's boundary conditions rather than indicators of genuine age precision. By retaining a higher density threshold of \(> 3\times10^9\) for most applications, we effectively exclude these edge stars, thereby ensuring that the majority of age estimates fall within a reliable uncertainty range of approximately 2 Gyr. This selection criterion strikes a balance between sample size and estimation accuracy, maintaining the robustness and reliability of our dataset.

    In Appendix \ref{sec:thresholdChange} we present how using different training density thresholds affects the various plots and maps presented in this work. Figure \ref{fig:agemap} specifically is repeated with different density thresholds in Appendix \ref{sec:agefigure} to illustrate the effects of both allowing the model to extrapolate ages and constraining the model to only the parameter space well sampled by the training set.


    
        \subsection{Estimating C \& N Abundances with Normalizing Flows}\label{CandN}

\begin{figure}[ht!]
                \centering
                \includegraphics[width=0.9\linewidth]{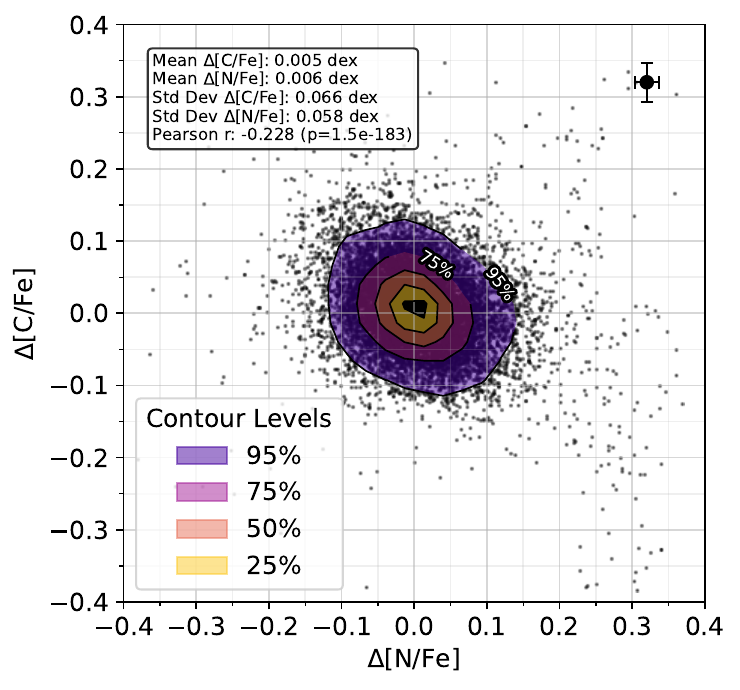}
                \caption{Scatter plot showing the differences between the model-estimated carbon and nitrogen abundances and the measured abundances from APOGEE. Contour lines indicate density levels, representing the percentage of data points enclosed within each contour. A representative error bar in the upper right corner illustrates the typical uncertainties in the measured abundances. Most stars have estimated C \& N abundances within 0.1 dex of the measured values, corresponding to an age uncertainty of approximately 2 Gyr. Additional statistics are displayed in the upper left corner.}
                \label{fig:CAndNDiff}
            \end{figure}

            Using the process described in Section \ref{2dCPf} we calculated the 2D C \& N probability distributions for each star in the training and test sets. We evaluated the model's overall ability to estimate the stellar C \& N abundances by comparing the difference between the peak probabilities and the real  C \& N abundances. These comparisons are shown in Figure \ref{fig:CAndNDiff}. Most stars in the test set have estimated C \& N abundances within \(< 0.1\) dex of the real values. The Pearson correlation coefficient between the differences in [C/Fe] and [N/Fe] is -0.22), indicating a weak negative correlation. This slight correlation corresponds with changes in [C/N], which is related to stellar age. An opposite correlation indicates lines of constant [C/N].

            A 0.1 dex spread in [C/N] corresponds with an age spread of \(\approx 2\) Gyr, which is consistent with the age uncertainties from the model and our other validation analyses.  Figure \ref{fig:CAndNDiff} also reveals some interesting structure among the outliers. Particularly, there is a string of stars with a real [C/N] much lower than estimated by the model. These stars are a mix of young (\(\leq 2 \) Gyr)  stars and stars that sit below the main [C/N] - Age relation shown in Figure \ref{fig:ParamSpace} (right). Another set of outliers in Figure \ref{fig:CAndNDiff} consists of stars with a higher C \& N than estimated. We identified these stars as having an unusually high \(C + N\) given their [Fe/H]. Notably, these stars sit on a constant [C/N] diagonal that goes through the origin, indicating that the model did estimate the correct [C/N] ratio for these stars.

    \subsection{Chemo-Age Cartography}   \label{sec:cartography} 
        
\begin{figure*}[ht!]
            \centering
            \includegraphics[width=1\linewidth]{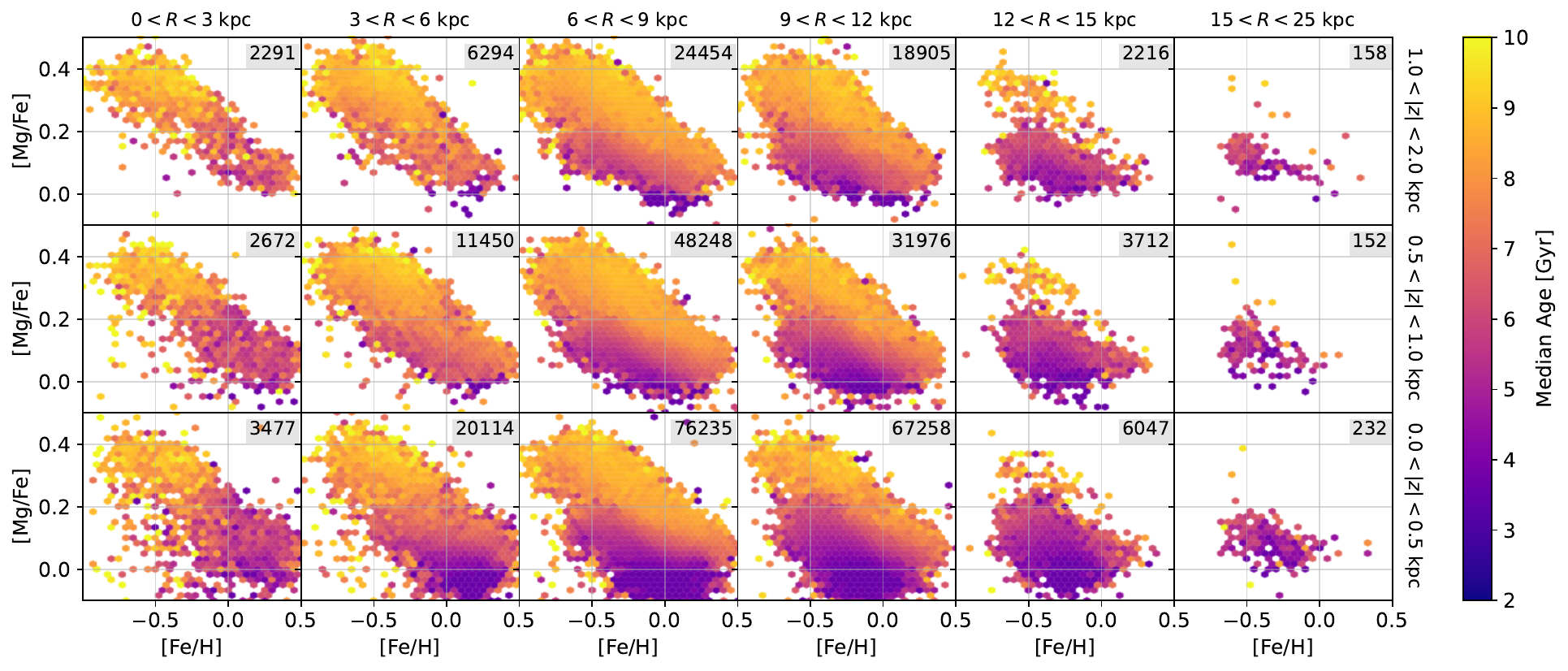}
            \caption{Hexbin map of stars in the in the [Mg/Fe] vs. [Fe/H] plane as a function of R and \(|Z|\), color coded by median age, produced similarly to \citet[Figure~9]{Imig_2023}. The number in the top-right corner of each panel is the number of stars in that spatial bin we have ages for, using a training density threshold of \(3\times10^9\). This figure is repeated in Appendix \ref{sec:chemofigure} with different density thresholds to illustrate how changes to the threshold affect the map.}
            \label{fig:Chemo-age}
        \end{figure*}

        In Figure \ref{fig:Chemo-age} we present a hexbin map of the spatial distribution of [Mg/Fe] coded by median stellar age from our model. This figure was produced similarly to \citet[Figure~9]{Imig_2023} in order to highlight the capabilities of our model.  We see much of the same structure as seen in \citet{Imig_2023}: a high-\(\alpha\) sequence composed of older stars across all radii, and a low-\(\alpha\) sequence that has more age variation across a radial gradient.
        Our new catalog contains ages for stars with \(\rm{[Fe/H]\le-0.7}\) which were excluded in \citet{Imig_2023} due to stricter limitations in the age model from \citet{DistMass}. Figure \ref{fig:Chemo-age} also includes the number of stars included in each region. In general, with the new age catalog, we have stellar ages for more stars in most regions, and in bins around the solar neighborhood, we have an order of magnitude more stars.

\section{Conclusion}
In this work, we have demonstrated the effectiveness of normalizing flows for estimating stellar ages with robust uncertainty quantification, applied to the extensive dataset provided by SDSS-V Milky Way Mapper. By leveraging the flexibility of normalizing flows, we have produced the largest catalog of ages for evolved stars, encompassing 378,72 stars when using our recommended training density threshold. Our methodology accounts for both observational and training-data-driven uncertainties by learning the joint distribution of stellar parameters and their associated errors, ensuring that the derived age estimates are both accurate and representative of the underlying data distribution.

While our model performs well within well-sampled regions of parameter space, challenges remain in extrapolating to less densely represented areas, particularly for metal-poor (\(\rm{[Fe/H] <-1}\)) and upper RGB stars (\(\log g < 1\)). However, the model’s performance in other regions demonstrates its general robustness and adaptability.

All of our derived ages, uncertainties, training space density, and the full posteriors are available in a value added catalog (VAC) for SDSS MWM DR19. A data model is presented in Appendix \ref{sec:datamodel} 

The methods presented here pave the way for further improvements in spectroscopic age determination, particularly through hybrid model approaches that incorporate full spectra and multiple data modes, such as combining normalizing flows with variational autoencoders or integrating more types of data to enhance age precision.

Looking forward to applications of our ages, this stellar age catalog opens new opportunities for studying the chemical and dynamical evolution of the Galaxy. We presented an early piece of this work in Section \ref{sec:cartography} and will continue to explore novel applications in galactic archaeology, such as mapping age gradients across stellar streams and the Galactic halo. By the end of SDSS-V we will have access to millions more stellar spectra across parameter space, many of which will be evolved stars that we can estimate ages for. The future potential for exploring chemo-dynamical-age relations across the Galaxy is substantial, with our methods providing valuable tools for future SDSS data releases and ongoing studies in Galactic archaeology. This work represents an important step toward more precise age estimates that will support a deeper understanding of Galactic structure and evolution.


\section*{Acknowledgments}

We are grateful to Juna Kollmeier, Jamie Tayar, and José G. Fernández-Trincado for valuable discussions and feedback. Much of the discussion in this manuscript was inspired by conversations with various conference and seminar attendees, including Phil Van-Lane, James Johnson, Jennifer Johnson, Marc Pinsonneault, Jason Jackiewicz, David Weinberg, Phillip Cargile, Henry Leung, Joshua Speagle, and Paul Beck.

A.S-M. gratefully acknowledges support from SDSS-V.

E.J.G. is supported by an NSF Astronomy and Astrophysics Postdoctoral Fellowship under award AST-2202135

Funding for the Sloan Digital Sky Survey V has been provided by the Alfred P. Sloan Foundation, the Heising-Simons Foundation, the National Science Foundation, and the Participating Institutions. SDSS acknowledges support and resources from the Center for High-Performance Computing at the University of Utah. SDSS telescopes are located at Apache Point Observatory, funded by the Astrophysical Research Consortium and operated by New Mexico State University, and at Las Campanas Observatory, operated by the Carnegie Institution for Science. The SDSS web site is \url{www.sdss.org}.

SDSS is managed by the Astrophysical Research Consortium for the Participating Institutions of the SDSS Collaboration, including the Carnegie Institution for Science, Chilean National Time Allocation Committee (CNTAC) ratified researchers, Caltech, the Gotham Participation Group, Harvard University, Heidelberg University, The Flatiron Institute, The Johns Hopkins University, L'Ecole polytechnique f\'{e}d\'{e}rale de Lausanne (EPFL), Leibniz-Institut f\"{u}r Astrophysik Potsdam (AIP), Max-Planck-Institut f\"{u}r Astronomie (MPIA Heidelberg), Max-Planck-Institut f\"{u}r Extraterrestrische Physik (MPE), Nanjing University, National Astronomical Observatories of China (NAOC), New Mexico State University, The Ohio State University, Pennsylvania State University, Smithsonian Astrophysical Observatory, Space Telescope Science Institute (STScI), the Stellar Astrophysics Participation Group, Universidad Nacional Aut\'{o}noma de M\'{e}xico, University of Arizona, University of Colorado Boulder, University of Illinois at Urbana-Champaign, University of Toronto, University of Utah, University of Virginia, Yale University, and Yunnan University.

This research was supported by the Munich Institute for Astro-, Particle and BioPhysics (MIAPbP) which is funded by the Deutsche Forschungsgemeinschaft (DFG, German Research Foundation) under Germany´s Excellence Strategy – EXC-2094 – 390783311.

\appendix
\section{StarFlow Catalog Data Model}\label{sec:datamodel}
Our value‐added catalog consists of three files, each row‐matched to SDSS-V MWM DR19. The primary file—recommended for most users—provides maximum likelihood estimates along with $\pm1\sigma$ error bars for both age and mass, as well as the training space density parameter for each star. The other two files contain the full posterior distributions for the age and mass models, respectively. Table \ref{tab:VAC} details the complete contents of the primary catalog file.

\begin{table}[hb!]
    \centering
    \begin{tabular}{|c|c|c|}
        \hline
         Data& Header Name & Description\\ \hline \hline
         SDSS-V ID& sdss\_id & Unique SDSS-V ID\\ \hline
         APOGEE ID/2MASS ID& sdss4\_apogee\_id & ID from 2MASS\\ \hline
         Stellar age&age& Maximum likelihood age from the StarFlow age model\\ \hline
         Positive age error&e\_p\_age  & $+1\sigma$ age uncertainty\\ \hline
         Negative age error& e\_n\_age & $-1\sigma$ age uncertainty\\ \hline
         Stellar mass& mass & Maximum likelihood mass from the StarFlow mass model\\ \hline
         Positive mass error& e\_p\_mass & $+1\sigma$ mass uncertainty\\ \hline
         Negative mass error& e\_n\_mass & $-1\sigma$ mass uncertainty\\ \hline 
         Training space density& training\_density & The density parameter described in Section \ref{sec:density}\\ \hline 
         BITMASK& BITMASK & Contains flags to indicate notes about a given star\\ \hline 
    \end{tabular}
    \caption{Overview of the StarFlow primary catalog files contents}
    \label{tab:VAC}
\end{table}

\section{Using a different training density threshold}\label{sec:thresholdChange}

Here we present how parameter space coverage (Figure \ref{fig:Density}), Galactic age map (Figure \ref{fig:agemap}), and chemo-age map (Figure \ref{fig:Chemo-age}) change when varying the training density threshold used.

 Lower density thresholds allow the model to extrapolate beyond the well-sampled regions, increasing coverage but also introducing the risk of unreliable age estimates in sparsely populated areas. In contrast, higher thresholds restrict extrapolation, enhancing reliability but reducing the number of stars with estimated ages. Our recommended density threshold is \(3\times10^9\).

\subsection{Parameter space coverage of stars for different training density thresholds}\label{sec:densityfigure}
\begin{figure}[ht!]
                \centering
                \gridline{\fig{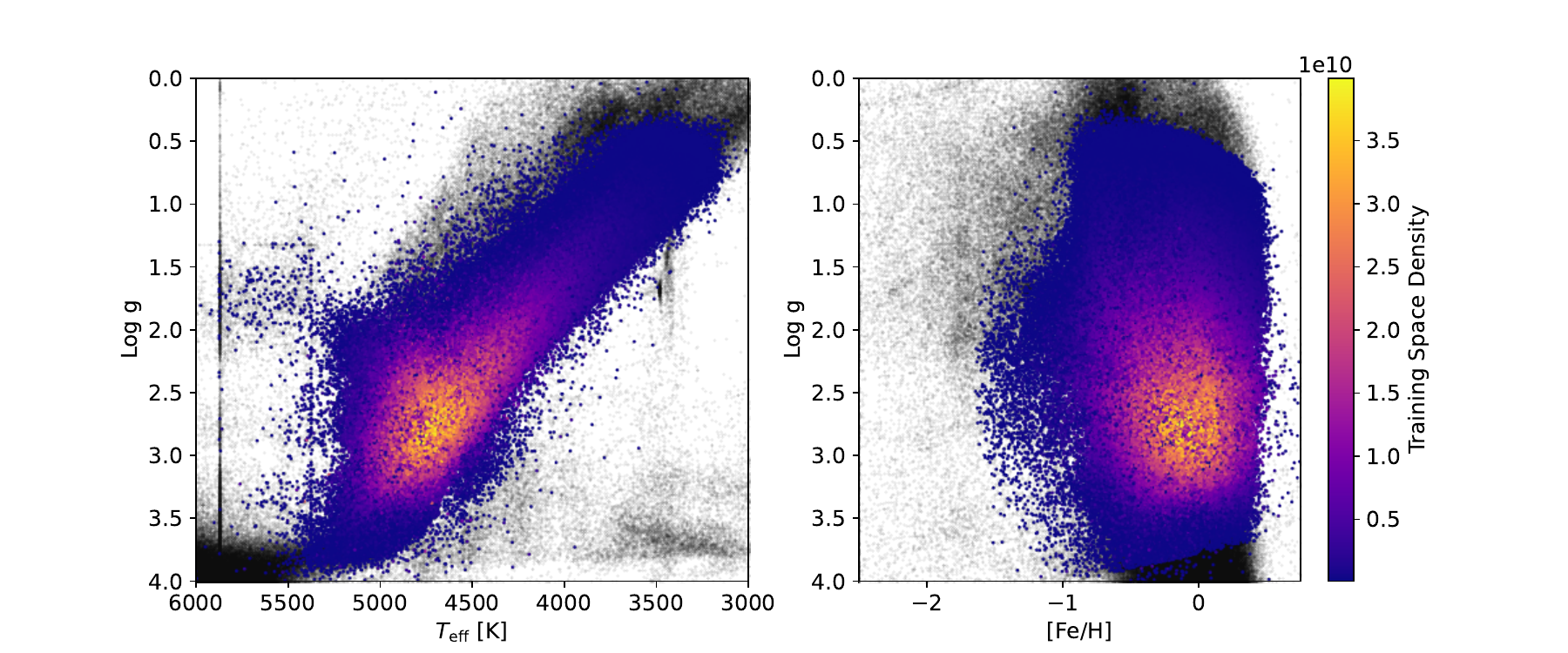}{0.85\textwidth}{}}
                \gridline{\fig{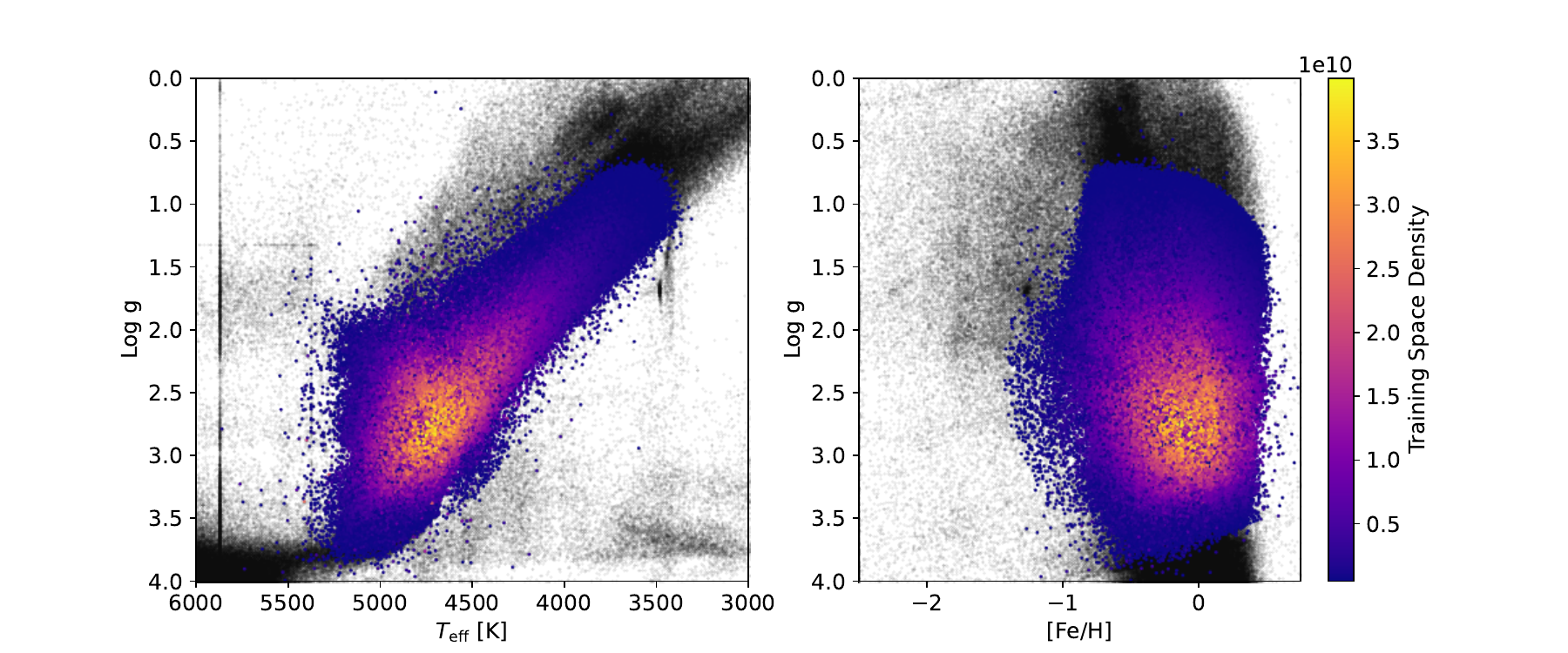}{0.85\textwidth}{}}                
                \gridline{\fig{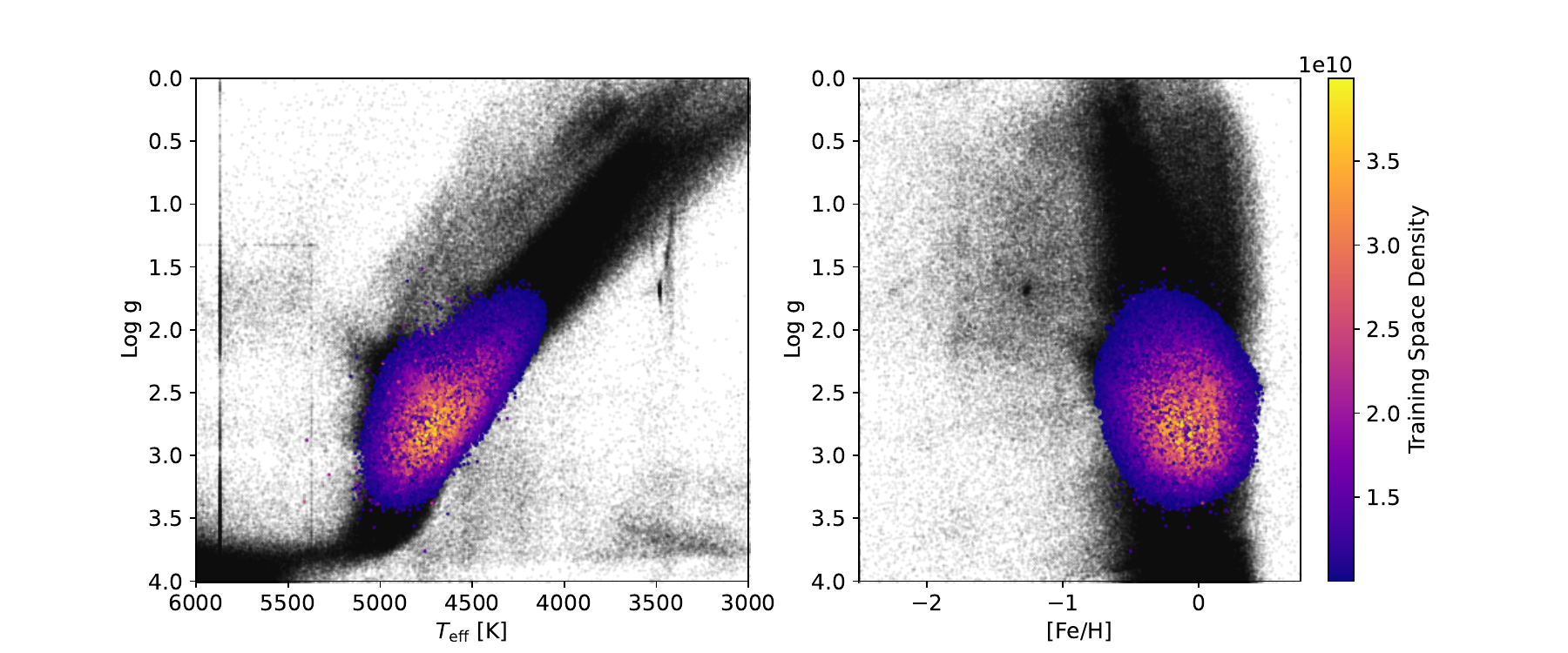}{0.85\textwidth}{}}                
                \caption{The parameter space coverage of the model at: \textbf{(Top)} Training space density \( > 10^8\). \textbf{(Center}) Training space density \( >5 \times 10^8\). \textbf{(Bottom}) Training space density \( >10^{10}\).}
                \label{fig:Density10_8}
            \end{figure}

Figure \ref{fig:Density10_8} show how changing the density threshold influences Figure \ref{fig:Density}. The first two panels use lower thresholds, which allows the model to extrapolate beyond the original training parameter space. The third panel uses a higher threshold, showing how the parameter space is constrained when using training density thresholds higher than our recommended start value.

    \subsection{Galactic Age Map using different training density thresholds}\label{sec:agefigure}

\begin{figure}[ht!]
        \centering
        \gridline{\fig{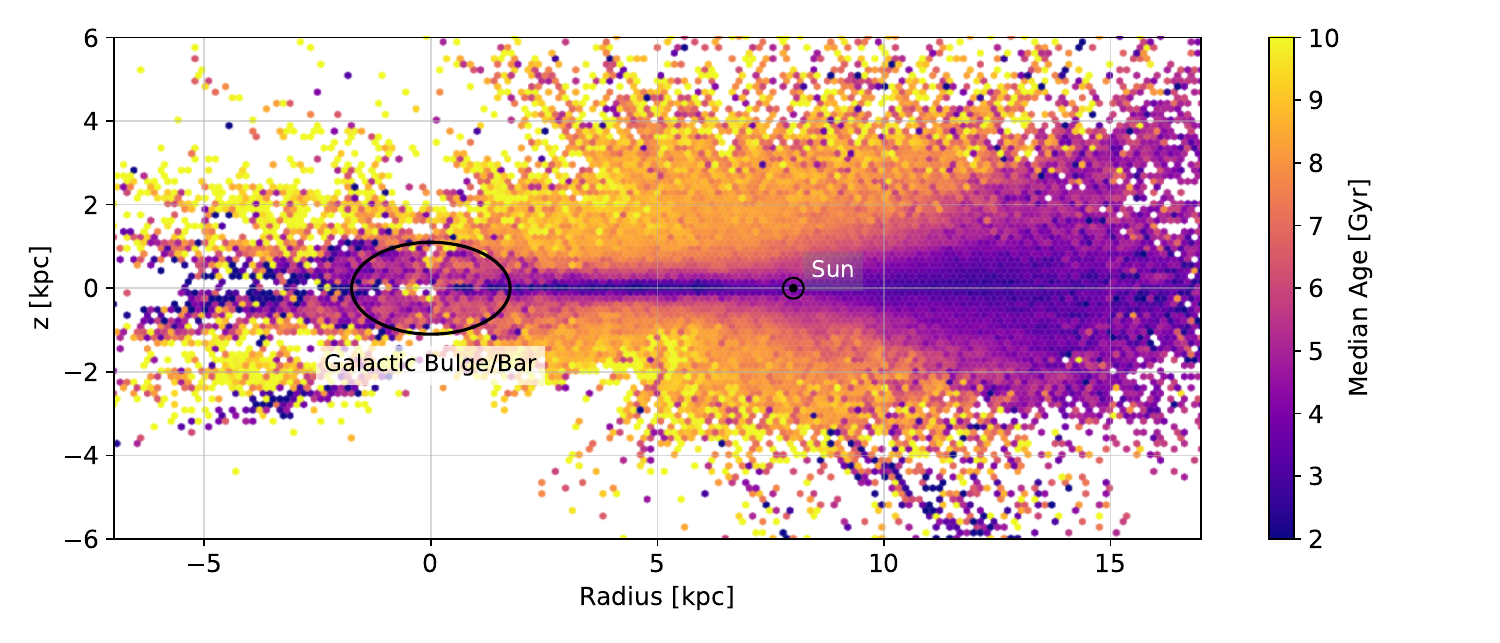}{0.75\textwidth}{}}
        \gridline{\fig{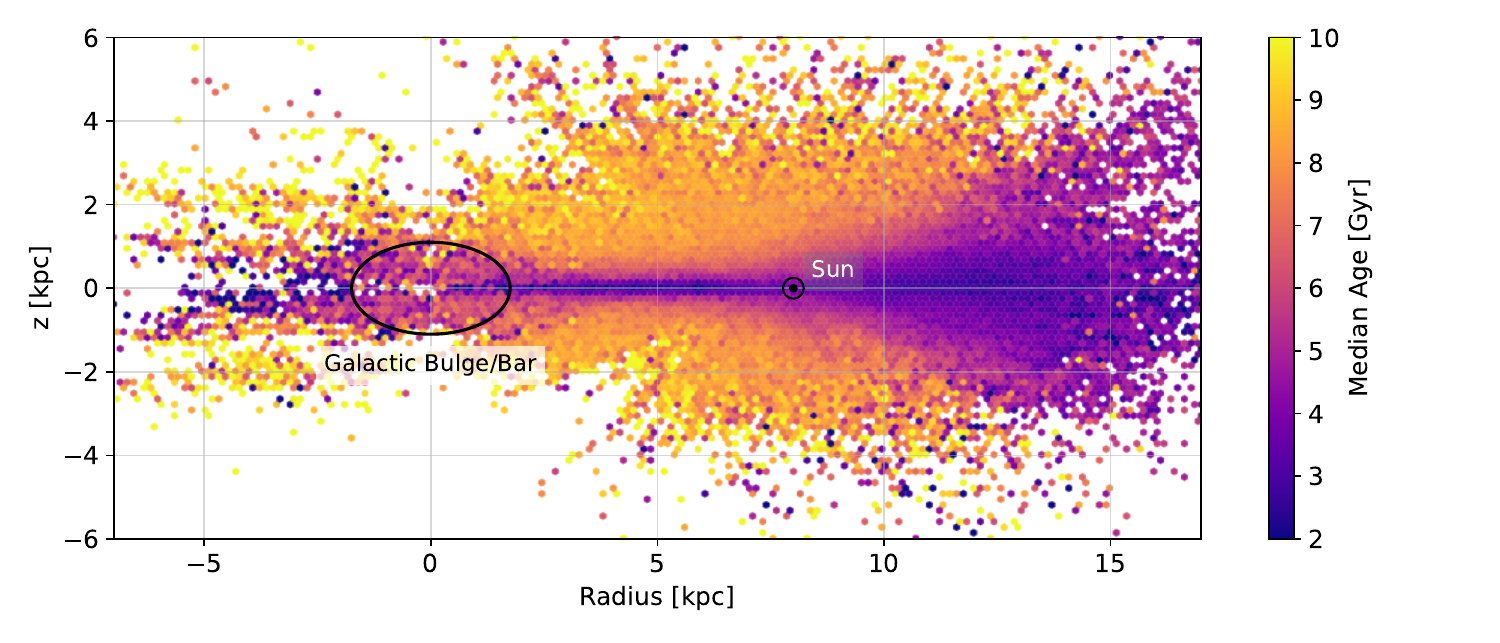}{0.75\textwidth}{}}                
        \gridline{\fig{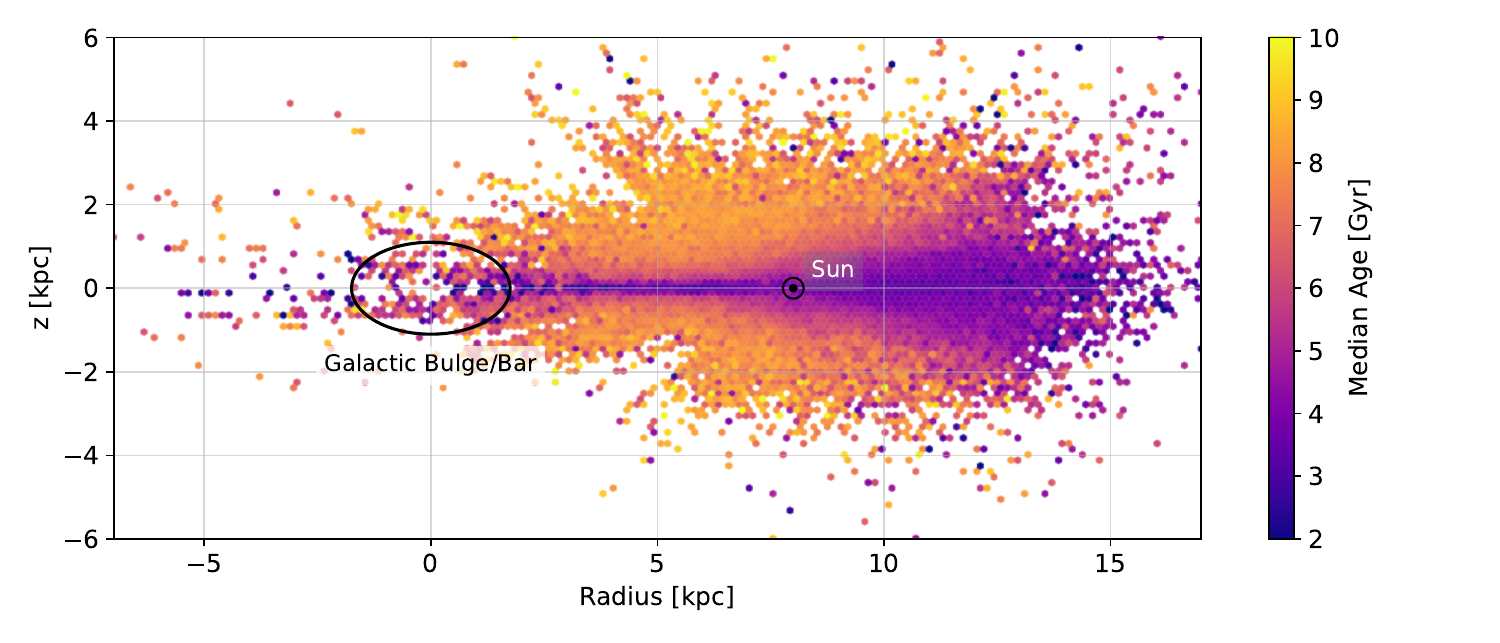}{0.75\textwidth}{}}                
        \caption{Galactic age maps using a threshold of:\textbf{(Top)} Training space density \( > 10^8\). \textbf{(Center}) Training space density \( >5 \times 10^8\). \textbf{(Bottom}) Training space density \( >10^{10}\).}
        \label{fig:agemap10_8}
    \end{figure}

    Figure \ref{fig:agemap10_8} shows how our Galactic stellar age map (Figure \ref{fig:agemap}) changes when using different training density thresholds. 
    Again, the upper two figures show lower thresholds and thus more age extrapolation. And the lower panel shows a tighter constraint.
    
    
    

    \subsection{Chemo-Age map using different training density thresholds}\label{sec:chemofigure}

\begin{figure}[ht!]
        \centering
        \gridline{\fig{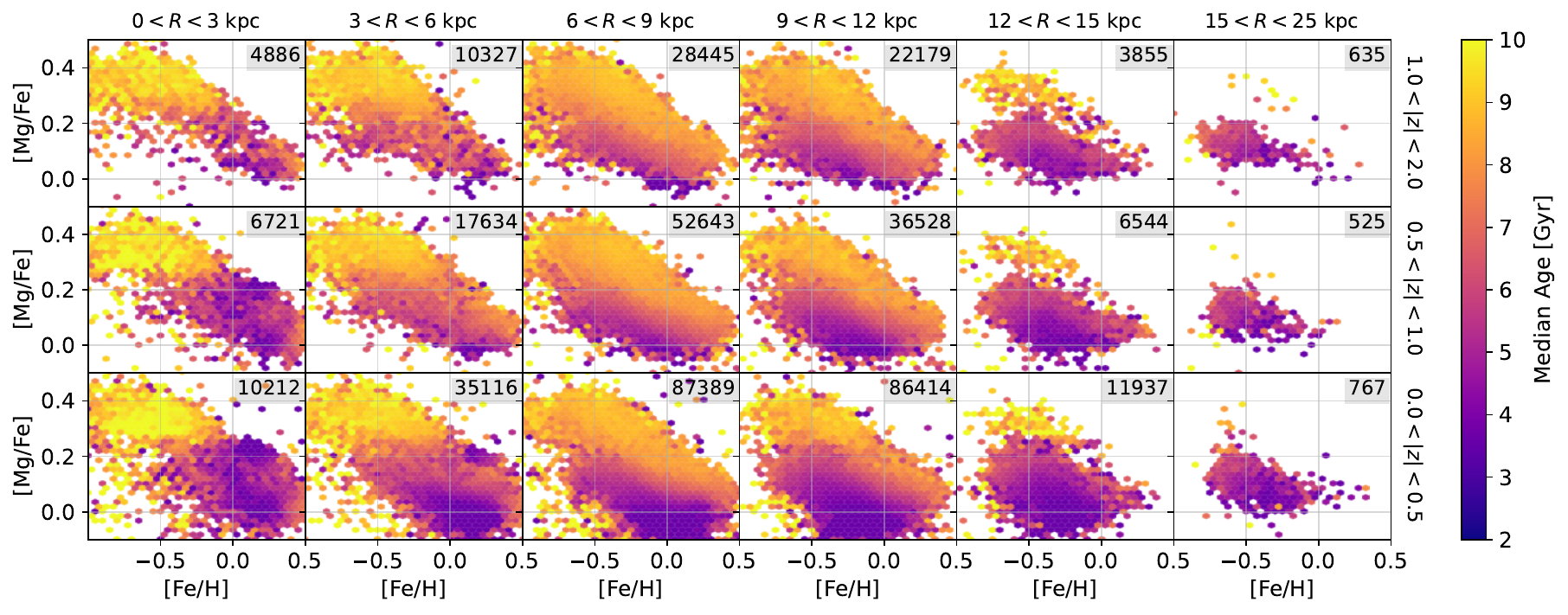}{0.75\textwidth}{}}
        \gridline{\fig{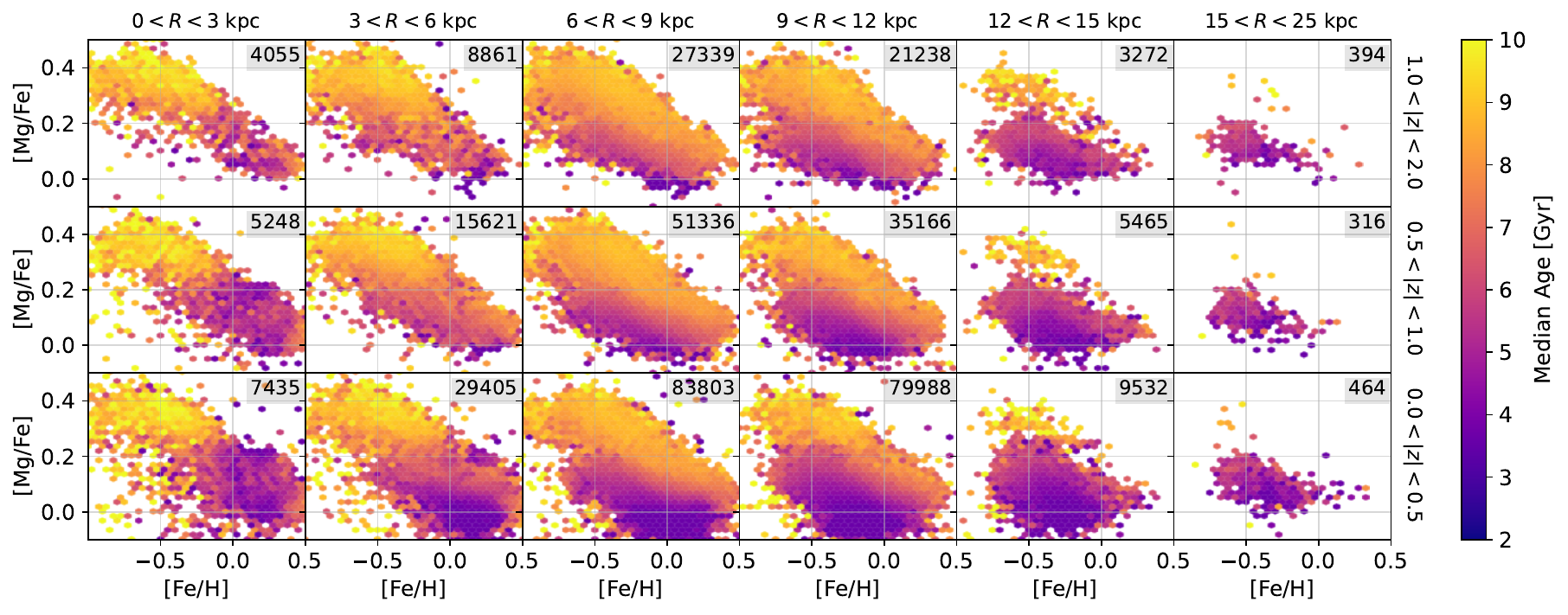}{0.75\textwidth}{}}                
        \gridline{\fig{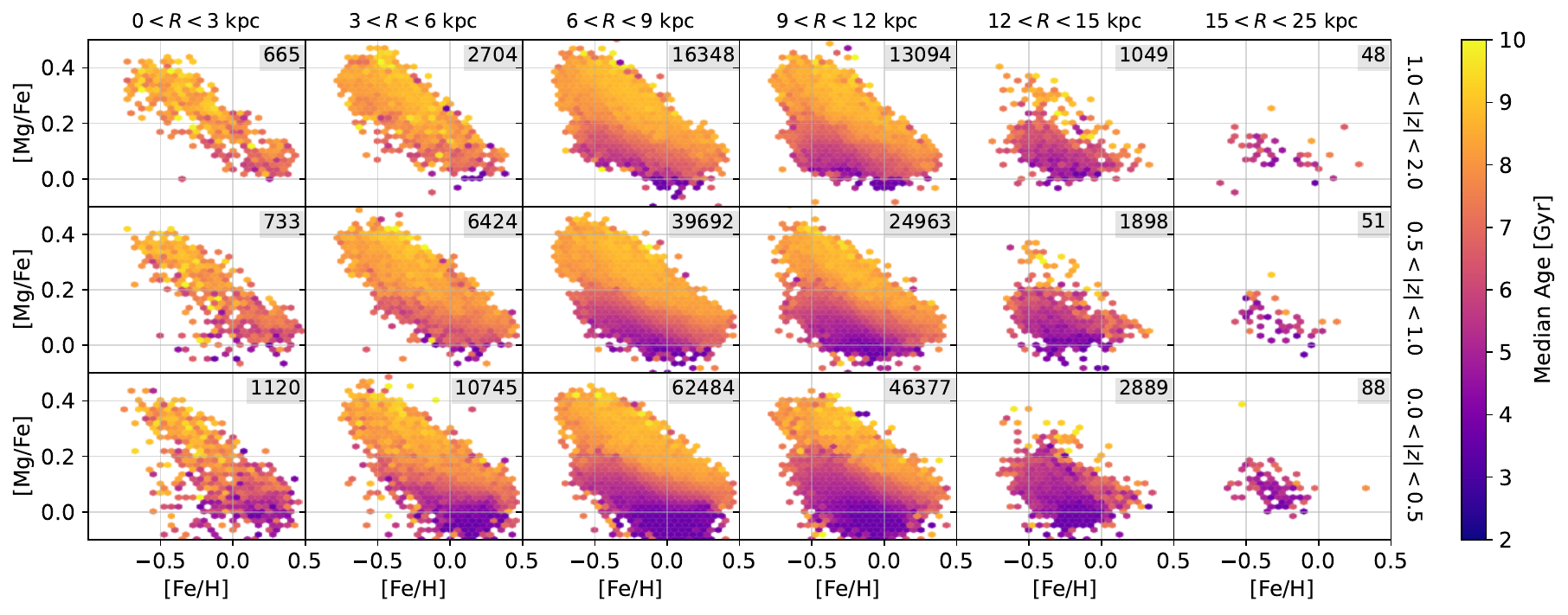}{0.75\textwidth}{}}                
        \caption{\(\alpha\)-Metallicity plane maps using a threshold of: \textbf{(Top)} Training space density \( > 10^8\). \textbf{(Center}) Training space density \( >5 \times 10^8\). \textbf{(Bottom}) Training space density \( >10^{10}\).}
        \label{fig:chemoagemap_8}
    \end{figure}

    Figure \ref{fig:chemoagemap_8} shows how our Chemo-Age map (Figure \ref{fig:Chemo-age}) changes when using different training density thresholds. 
    
    
    

    These results illustrate the trade-off between extrapolation and reliability when selecting a training density threshold. They provide context for the choice made in Section \ref{sec:density}. Users may choose to use a different density threshold than our recommended \(3\times10^9\), however we strongly do not recommend using a threshold any lower than \(10^8\). 

\section{Example posteriors}\label{sec:examples}
\begin{figure}[ht!]
        \centering
        \gridline{\fig{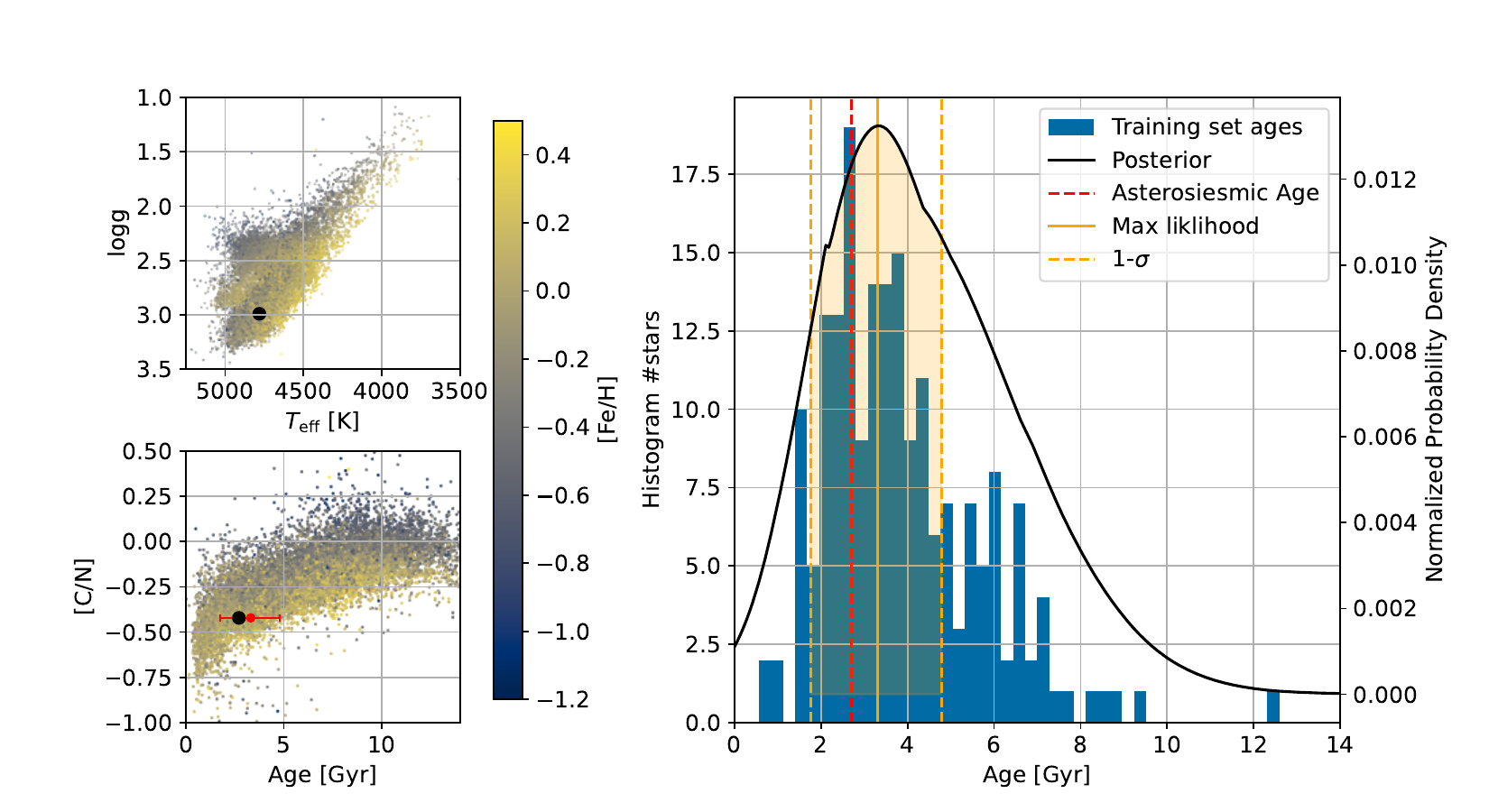}{0.6\textwidth}{}}
        \gridline{\fig{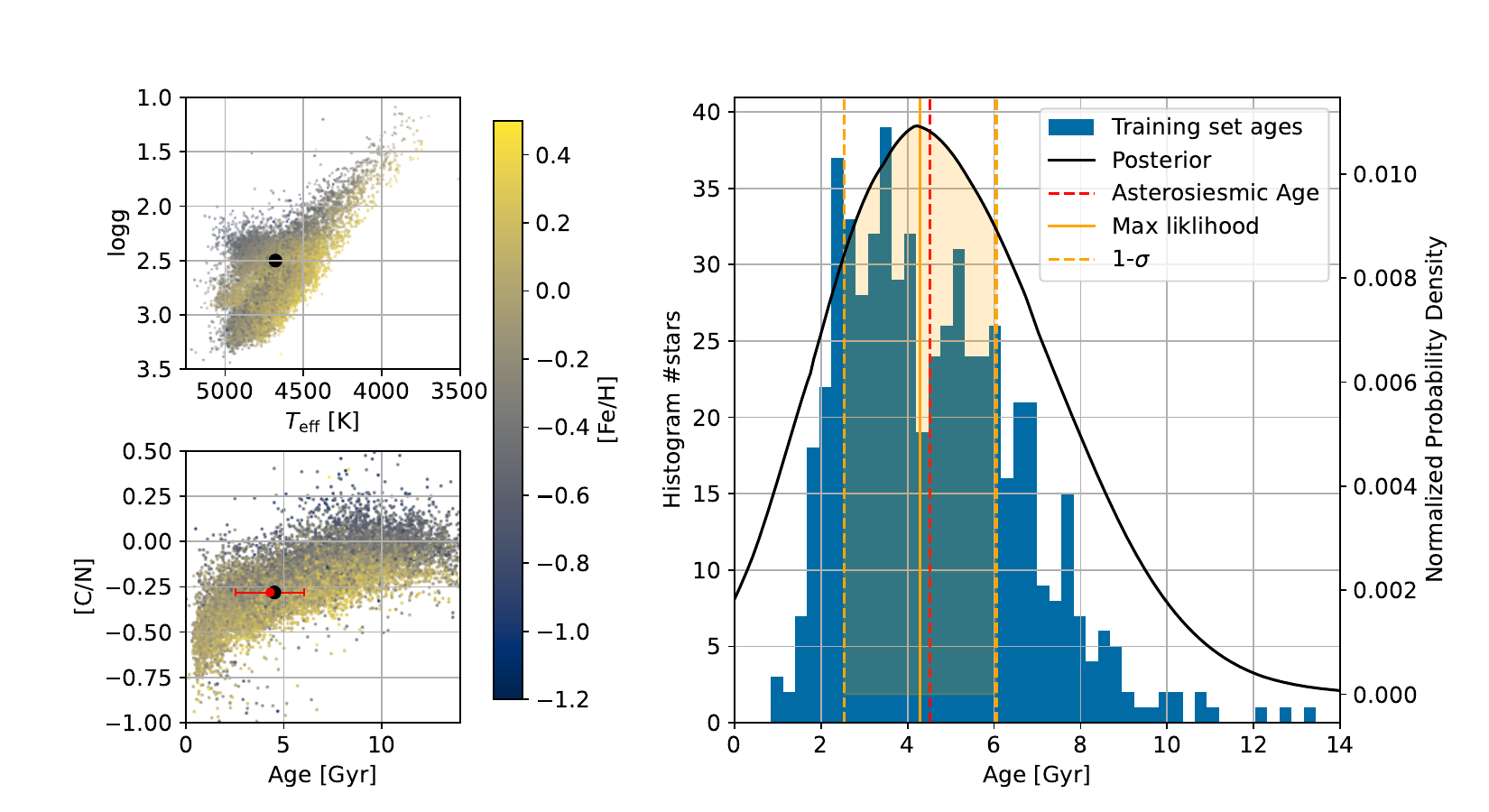}{0.6\textwidth}{}}                
        \gridline{\fig{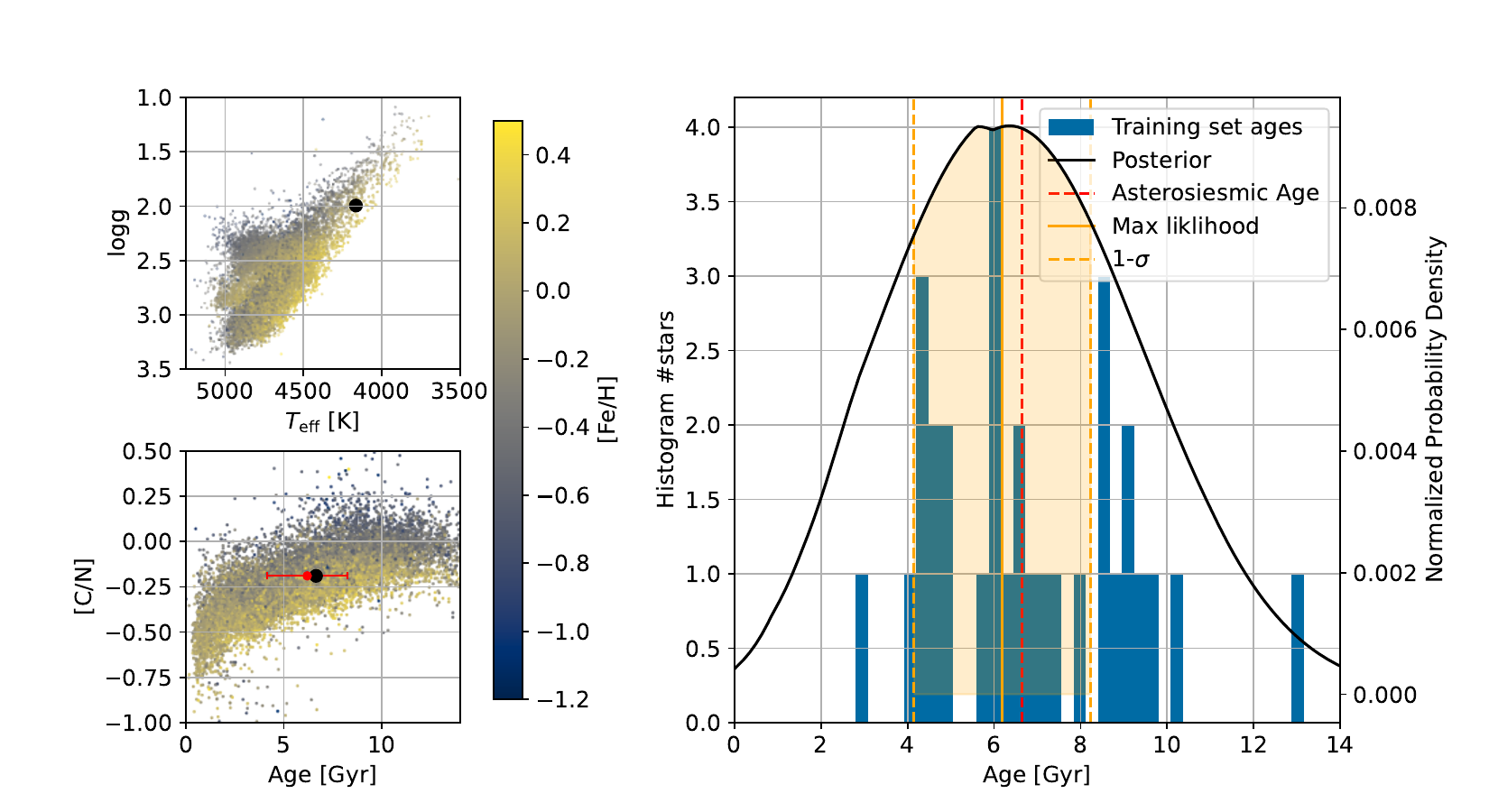}{0.6\textwidth}{}}                
        \caption{Additional examples of test-set age posterior recoveries using the same format as Figure 2. Each panel shows a test-set star with a different asteroseismic age: (Top) a young star ($\approx 2.5$ Gyr), (Middle) a slightly older star ($\approx4.5 Gyr$), and (Bottom) an middle-age star ($\approx6.5$ Gyr). These examples illustrate how the model recovers age posteriors with varying shape and uncertainty across the age range. Black curves show the full posterior, orange lines indicate the maximum likelihood age, and dashed lines mark the 1$\sigma$ range. Red dashed lines show the asteroseismic age.}
        \label{fig:examples}
    \end{figure}

To complement Figure \ref{fig:AgePosterior} in the main text, we include additional examples of test-set stars across a broader range of parameters and age shown in Figure \ref{fig:examples}.

    \clearpage

\bibliography{STARFLOW}{}
\bibliographystyle{aasjournal}



\end{document}